\documentclass[11pt,onecolumn,journal]{IEEEtran}
\usepackage[T1]{fontenc}
\usepackage{amsmath,amssymb,amsthm,mathtools,bm}
\usepackage{microtype}
\usepackage{hyperref}
\hypersetup{
colorlinks,
linkcolor=blue,
anchorcolor=blue,
citecolor=red
}

\usepackage{xcolor}
\usepackage[noadjust]{cite}
\usepackage{booktabs}
\newtheorem{theorem}{Theorem}
\newtheorem{lemma}[theorem]{Lemma}
\newtheorem{proposition}[theorem]{Proposition}

\newtheorem{definition}[theorem]{Definition}
\theoremstyle{remark}\newtheorem{remark}[theorem]{Remark}
\DeclareMathOperator{\Tr}{Tr}
\DeclareMathOperator{\supp}{supp}

\newcommand{\E}{\mathbb E}
\newcommand{\id}{\mathbf 1}
\newcommand{\norm}[1]{\left\|#1\right\|}

\allowdisplaybreaks[2]

\title{Exponential Analysis for Trace-Distance Errors \\in Quantum Information Theory}

\author{Hao-Chung~Cheng, Li~Gao, Shi-Bing~Li, Hongsen~Qiu%
\thanks{Hao-Chung Cheng is with the Department of Electrical Engineering and Graduate Institute of Communication Engineering, National Taiwan University, Taipei, Taiwan; the Department of Mathematics, National Taiwan University, Taipei, Taiwan; the Center for Quantum Science and Engineering, National Taiwan University, Taipei, Taiwan; the Physics and Mathematics Divisions, National Center for Theoretical Sciences, Taipei, Taiwan; and the Hon Hai (Foxconn) Quantum Computing Center, New Taipei City, Taiwan.}%
\thanks{Li~Gao is with the School of Mathematics and Statistics, Wuhan University, Wuhan, China; Wuhan Institute of Quantum Technology, Wuhan, China.}
\thanks{Shi-Bing Li is with the Insitute for Advanced Study in Mathematics, Harbin Insitute of Technology, Harbin, China.}
\thanks{Hongsen Qiu is with the Insitute for Advanced Study in Mathematics, Harbin Insitute of Technology, Harbin, China.}
}
\date{}
\begin{document}
\maketitle

\begin{abstract}
We study the trace-distance error in classical-quantum soft covering,
privacy amplification against quantum side information, convex splitting,
and decoupling by a Haar-random unitary followed by partial trace.
In the independent and identically distributed setting, we determine
exact reliability functions and strong converse exponents for soft
covering with independent codewords, uniform convex splitting, and
Haar partial-trace decoupling. For privacy amplification, we determine
the exact reliability function for independent random binning and the
exact strong converse exponent for every two-universal hash ensemble.
The reliability and strong converse formulas involve sandwiched and
Petz-R\'enyi information, respectively. The proofs connect trace-distance
errors to quantum hypothesis testing, using Rosenthal inequalities for
sums of random matrices in the reliability analysis and conditional
second-moment bounds for the strong converses.
\end{abstract}

%\begin{IEEEkeywords}
%Decoupling, error exponents, privacy amplification, soft covering, trace distance.
%\end{IEEEkeywords}

\section{Introduction}

An important class of problems in quantum information theory concerns the simulation of quantum states, correlations, and channels, often with the additional requirement of decoupling an output from its environment. In this work, We study the error exponents (also called reliability function) and strong converse exponents for four such \emph{covering-type
problems}~\cite{ChengGaoHircheHuangLiu2025Sharp}.

\begin{itemize}
\item \textbf{Classical-quantum (c-q) soft covering.}
Soft covering approximates the average output state of a c-q channel using
limited input randomness. A codeword to the channel input is selected uniformly from a codebook whose entries are sampled independently from the prescribed
input distribution \cite[Sec.~9.4]{Hayashi2017QuantumInformationTheory},
\cite{HayashiChengGao2024,ChengGao2024,ShenGaoCheng2024}. 
Soft covering is used in wiretap channel coding to make the eavesdropper’s output nearly independent of the transmitted message~\cite{CaiWinterYeung2004,Devetak2005,Hayashi2015}.
Quantum covering also plays a role in channel
identification~\cite{AhlswedeWinter2002Identification}.

\item \textbf{Privacy amplification against quantum side information.}
Privacy amplification extracts a nearly uniform key from a classical source while suppressing correlations with an adversary's quantum side information~\cite{TomamichelSchaffnerSmithRenner2011,shen2022strong,ShenGaoCheng2024,ShenGaoCheng2023RegularBinning,RegulaTomamichel2026Rethinking}. It is a central step in composably secure quantum key distribution~\cite{RennerKonig2005,Renner2005Thesis,PortmannRenner2022,TanRegulaTomamichel2026}.

\item \textbf{Convex splitting.}
Convex splitting uses random placement to dilute the correlations in
a bipartite state \(\rho_{AB}\)~\cite{ChengGao2023Convex}. One of the \(M\) registers is chosen
uniformly to hold the \(A\) subsystem, and each of the remaining
\(M-1\) registers is prepared independently in a reference state
\(\tau_A\). Averaging over the choice of register gives a joint state
that should approximate \(\tau_A^{\otimes M}\otimes\rho_B\), where
\(\rho_B\) is the input marginal on \(B\). 
Convex splitting was introduced for
the one-shot quantum state splitting protocol~\cite{AnshuDevabathiniJain2017, AnshuJainWarsi2018Redistribution} and measurement
compression with quantum side information~\cite{AnshuJainWarsi2019Measurement}.
Its multipartite extension gives protocols for quantum broadcast
channel simulation~\cite{BertaChengGao2025Multipartite}.

\item \textbf{Standard (partial-trace) decoupling.}
Decoupling seeks to make a quantum subsystem independent of its
environment. The standard method applies a Haar-random unitary
and discards part of the system. The retained subsystem is required to
be nearly maximally mixed. This procedure is used in fully quantum Slepian--Wolf
coding~\cite{AbeyesingheDevetakHaydenWinter2009} and in proofs of quantum
channel coding theorems~\cite{HaydenHorodeckiWinterYard2008,ChengDupuisGao2024}.
The decoupling method also underlies protocols for quantum state
merging~\cite{HorodeckiOppenheimWinter2007,DupuisBertaWullschlegerRenner2014,BertaChengYao2026}.
\end{itemize}
We measure errors of these four problems in trace distance, which is a natural distance that quantifies how well a
two-outcome quantum measurement distinguishes the target state and the processed state~\cite{Holevo1973Decision,Helstrom1976,RegulaTomamichel2026Rethinking}.

A key ingredient in the achievability arguments for all four tasks is the use of \emph{randomness}.
Earlier one-shot analyses exploit this randomness to derive error bounds that decay exponentially in the independent and identically distributed (i.i.d.) source.
For soft covering, Cheng and
Gao obtained an achievable trace-distance exponent for independently
sampled codebooks~\cite{ChengGao2024}. For privacy amplification, Dupuis
derived an exponent in terms of sandwiched R\'enyi conditional
entropy~\cite{Dupuis2023} (see also \cite{RegulaTomamichel2026Rethinking}). Analogous bounds are known for convex
splitting and
decoupling~\cite{ChengGao2023Convex,Dupuis2023,ChengDupuisGao2024}. A common mathematical tools in deriving these error exponent bounds are the method of complex interpolation.
%Throughout, we average the trace distance over realizations of the codebook, hash, or unitary.

Despite this progress, a natural question remained open: do these bounds give the sharp exponential decay rates of the trace-distance errors?
In classical soft covering, Yagli
and Cuff used Poissonization to prove a matching ensemble converse
for the expected total variation distance~\cite{YagliCuff2019}.
Yassaee further improved the polynomial prefactors~\cite{Yassaee2019SoftCovering}. For classical privacy
amplification, Hayashi derived an achievable total variation exponent
with side information, and Hayashi and Watanabe used fully independent
hash functions to obtain a matching bound above a critical extraction
rate~\cite{Hayashi2013TightPrivacy,HayashiWatanabe2016}.
It was unclear, however, how to extend these classical arguments to obtain matching trace-distance bounds to the quantum setting.
For c-q soft
covering and privacy amplification against quantum side information,  exact reliability functions are known under error criterion of sandwiched R\'enyi
divergence~\cite{LiQiuZhang2026Reliability}. Corresponding results for
privacy amplification under purified distance and relative entropy hold
at sufficiently high extraction rates~\cite{LiYaoHayashi2023}, while
relative-entropy exponents for partial-trace decoupling have been
determined at low discard rates~\cite{BertaChengYao2026}.

In this paper, we close the gap by proving matching converse bounds of trace distance errors for soft covering with independent codewords, privacy amplification by uniform random binning, uniform convex splitting, and Haar partial-trace decoupling. Together with the known achievability bounds, these converses determine the \emph{exact} reliability function \(\displaystyle\lim_{n\to\infty}-\frac{1}{n}\log\Delta_n\) for each of the four tasks. Here \(\Delta_n\) denotes the trace-distance error at blocklength \(n\), averaged over the random codebook, hash function, or Haar unitary when applicable. We further exhibit a strongly universal\(_2\) hash ensemble whose reliability exponent strictly exceeds that of independent uniform random binning.

Our proofs combine quantum hypothesis testing~\cite{hayashi2016correlation,MosonyiOgawa2015}
with functional-analytic tools, including noncommutative Rosenthal
inequalities~\cite{JungeXu2008RosenthalII} and H\"older's inequality.
For the reliability functions, these tools yield lower bounds on the
trace-distance errors and hence matching upper bounds on their
exponential decay rates. For the strong converse exponents, conditional
second-moment estimates allow us to lower-bound \(1-\Delta_n\) by a
constant multiple of a weighted binary hypothesis-testing cost.
This cost is the minimum weighted sum of type-I and type-II errors,
whose asymptotic exponent is characterized by Petz--R\'enyi
divergences~\cite{HiaiMosonyiOgawa}. The comparison therefore gives
matching upper bounds on the strong converse exponents.

We thereby establish the tightness of previously obtained exponential
strong converse bounds for the tasks considered
here~\cite{ChengGao2024,shen2022strong,SalzmannDatta2022,ChengGao2023Convex,ChengDupuisGao2024}.
Specifically, we determine
\(\displaystyle \lim_{n\to\infty}-\frac{1}{n}\log(1-\Delta_n)\),
the exponential rate at which the trace-distance error, averaged over
the prescribed randomness when applicable, approaches one in the
strong converse regime. For privacy amplification, the formula holds
for every sequence of universal\(_2\) hash ensembles at the prescribed
asymptotic extraction rate. The resulting formulas are summarized in
Table~\ref{tab:intro-exponents}.

Table~\ref{tab:intro-exponents} collects the established reliability functions and strong converse exponents. Related exact strong converse results have been established for
classical soft covering optimized over
codes~\cite{HePradhanWinter2026}, for c-q soft covering and privacy
amplification with sandwiched R\'enyi divergence as the error
criterion~\cite{LiQiuZhang2026StrongConverse}, and for optimized quantum
privacy amplification under a composable fidelity
criterion~\cite{RubboliTomamichel2026}.

\begin{table}[ht]
\centering
\caption[Trace-distance exponents for the four tasks]{%
Trace-distance exponents for the four tasks.\par\smallskip
}
\label{tab:intro-exponents}
\small
\renewcommand{\arraystretch}{1.35}
\setlength{\tabcolsep}{7pt}
\begin{tabular*}{\linewidth}{@{}l@{\extracolsep{\fill}}ll@{}}
\toprule
\textbf{Task} & \textbf{Reliability function} & \textbf{Strong converse exponent} \\
\midrule
\begin{tabular}[c]{@{}l@{}}
Classical-quantum soft covering\\[-1pt]
{\footnotesize(Theorem~\ref{thm:soft-exponents})}
\end{tabular}
& \(\displaystyle\max_{1\leq\alpha\leq2}\frac{\alpha-1}{\alpha}
  \bigl(R-I_\alpha^*(X:E)_\rho\bigr)\)
& \(\displaystyle\max_{0\leq s\leq1}(1-s)
  \bigl(I_s(X:E)_\rho-R\bigr)\) \\
\addlinespace[3pt]
\begin{tabular}[c]{@{}l@{}}
Privacy amplification\\[-1pt]
{\footnotesize(Theorem~\ref{thm:pa-exponents})}
\end{tabular}
& \(\displaystyle\max_{1\leq\alpha\leq2}\frac{\alpha-1}{\alpha}
  \bigl(H_\alpha^*(X|E)_\rho-R\bigr)\)
& \(\displaystyle\max_{0\leq s\leq1}(1-s)
  \bigl(R-H_s(X|E)_\rho\bigr)\) \\
\addlinespace[3pt]
\begin{tabular}[c]{@{}l@{}}
Convex splitting\\[-1pt]
{\footnotesize(Theorem~\ref{thm:convex-exponents})}
\end{tabular}
& \(\displaystyle\max_{1\leq\alpha\leq2}\frac{\alpha-1}{\alpha}
  \bigl(R-D_\alpha^*(\rho_{AB}\|\tau_A)\bigr)\)
& \(\displaystyle\max_{0\leq s\leq1}(1-s)
  \bigl(D_s(\rho_{AB}\|\tau_A)-R\bigr)\) \\
\addlinespace[3pt]
\begin{tabular}[c]{@{}l@{}}
Haar partial-trace decoupling\\[-1pt]
{\footnotesize(Theorem~\ref{thm:dec:exact})}
\end{tabular}
& \(\displaystyle\max_{1\leq\alpha\leq2}\frac{\alpha-1}{\alpha}
  \bigl(H_\alpha^*(A|E)_\rho-\log(c/k)\bigr)\)
& \(\displaystyle\max_{0\leq s\leq1}(1-s)
  \bigl(\log(c/k)-H_s(A|E)_\rho\bigr)\) \\
\bottomrule
\end{tabular*}\vspace{3pt}
{\normalfont\raggedright
In the first three rows, \(\displaystyle R=\lim_{n\to\infty}\tfrac{1}{n}\log M_n\).
For soft covering and privacy amplification, \(M_n\) is the codebook
size and the key alphabet size, respectively. For convex splitting,
\(M_n\) is the total number of registers. One is chosen uniformly to
hold the input subsystem \(A^n\). For Haar partial-trace decoupling,
\(A=C\otimes D\), with retained dimension \(c=|C|\geq2\) and discarded
dimension \(k=|D|\). Here \(X\) is classical; \(\rho_{XE}\) is the input
in the first two rows, \(\rho_{AB}\) is the convex-splitting input,
\(\rho_{AE}\) for the decoupling, and \(\tau_A\) is the
convex-splitting reference state. The starred
quantities \(I_\alpha^*, H_\alpha^*, D_\alpha^*\) use sandwiched
R\'enyi divergence optimized over the second reference state,
whereas the quantities \(I_s,H_s,D_s\) use Petz-R\'enyi divergence with
the actual marginal on the second system. For privacy amplification,
the reliability formula is exact for independent random binning
with \(M_n\geq2\) for all sufficiently large \(n\). \par}
\end{table}

The rest of paper is organized as follows. Section~\ref{sec:preliminaries} develops the key testing and moment estimates
and concludes with an overview of the common proof strategy.
The
reliability functions and converse exponents for the four tasks above are established in Sections~\ref{sec:soft-covering}--\ref{sec:partial-trace} respectively.

\section{Preliminaries}
\label{sec:preliminaries}

\subsection{Notation}
A quantum system and its Hilbert space are denoted by the same letter,
such as $A$, and $|A|:=\dim A$. 
The spaces of linear and positive semidefinite operators on $A$ are denoted by
$\mathcal B(A)$ and $\mathcal{P}(A)$, respectively, and
\[
\mathcal S(A):=\{\rho\in\mathcal B(A):\rho\geq0,\ \Tr\rho=1\}
\]
is the set of states on $A$. The identity operator is $\id_A$ and
$\pi_A:=\frac{\id_A}{|A|}$ is the maximally mixed state.
Operators and maps acting on specified registers are understood to act
as the identity on the other registers when these are omitted.

For an operator $X$, let $X^*$ be its adjoint and
$|X|:=(X^*X)^{1/2}$. The Schatten norms are
\[
\|X\|_p:=(\Tr|X|^p)^{1/p},\qquad 1\leq p<\infty,
\]
and $\|X\|_\infty$ is the operator norm. 
For any two states $\rho$ and $\sigma$, the trace distance between $\rho$ and $\sigma$ is 
$\frac12\|\rho-\sigma\|_1$.
For a positive operator $X$, $\supp X$ denotes its support, and
$X^0$ is the projection onto that support. Negative powers are taken
on the support and extended by zero on its orthogonal complement.
For self-adjoint $X$, $\{X>0\}$ denotes its strictly positive spectral
projection. A quantum channel is a completely positive,
trace-preserving linear map.

Classical alphabets are denoted by calligraphic letters such as
$\mathcal X$, and $X$ also denotes the corresponding classical register
with orthonormal basis $\{|x\rangle:x\in\mathcal X\}$.
A classical--quantum state has the form
\[
\rho_{XE}=\sum_{x\in\mathcal X}P_X(x)|x\rangle\langle x|
\otimes\rho_E^x.
\]

Throughout the paper, we consider the four tasks with finite alphabets and finite-dimensional quantum systems. We write $[u]_+:=\max\{u,0\}$ and adopt the convention
$-\log 0:=+\infty$.

Let $\rho$ be a state and let $\sigma\geq0$ be positive operator on the same space. The relative entropy between $\rho$ and $\sigma$ is
\[
D(\rho\|\sigma):=
\begin{cases}
\Tr\rho(\log\rho-\log\sigma),
& \supp\rho\subseteq\supp\sigma,\\[1mm]
+\infty,
& \text{otherwise}.
\end{cases}
\]
Here and in the following, all logarithms are natural.
For $\alpha>1$, the sandwiched R\'enyi divergence between $\rho$ and $\sigma$ is
\begin{equation*}
D_\alpha^*(\rho\|\sigma):=
\begin{cases}
\displaystyle
\frac{1}{\alpha-1}\log\Tr
\left(
\sigma^{\frac{1-\alpha}{2\alpha}}
\rho\,
\sigma^{\frac{1-\alpha}{2\alpha}}
\right)^\alpha,
& \supp\rho\subseteq\supp\sigma,\\[3mm]
+\infty,
& \text{otherwise}.
\end{cases}
\end{equation*}
For $0<s<1$, the Petz-R\'enyi divergence between $\rho$ and $\sigma$ is
\begin{equation*}
D_s(\rho\|\sigma):=
\begin{cases}
\displaystyle
\frac{1}{s-1}\log\Tr\rho^s\sigma^{1-s},
& \Tr\rho^s\sigma^{1-s}>0,\\[3mm]
+\infty,
& \Tr\rho^s\sigma^{1-s}=0.
\end{cases}
\end{equation*}
The endpoint conventions are
\begin{equation*}
D_0(\rho\|\sigma):=-\log\Tr(\Pi_\rho\sigma),
\qquad
D_1(\rho\|\sigma)=D_1^*(\rho\|\sigma):=D(\rho\|\sigma).
\end{equation*}
Here $\Pi_\rho$ denotes the orthogonal projector onto
$\supp\rho$. These agree with the respective one-sided limits. Under the support
inclusion above, the limits at order one are finite.

Let $\rho_{AB}\in\mathcal S(AB)$, and fix
$\tau_A\in\mathcal S(A)$ satisfying
$\supp\rho_A\subseteq\supp\tau_A$.
Then
\[
\supp\rho_{AB}
\subseteq\supp\rho_A\otimes\supp\rho_B
\subseteq\supp\tau_A\otimes\supp\rho_B.
\]
We use the following two fixed-reference information quantities:
\begin{align*}
D_\alpha^*(\rho_{AB}\|\tau_A)
&:=\inf_{\eta_B\in\mathcal S(B)}
D_\alpha^*(\rho_{AB}\|\tau_A\otimes\eta_B),
&&\alpha>1,
\\
D_s(\rho_{AB}\|\tau_A)
&:=D_s(\rho_{AB}\|\tau_A\otimes\rho_B),
&&0\leq s<1.
\end{align*}
Here $D_\alpha^*(\rho_{AB}\|\tau_A)$ optimizes over the reference state on $B$,
and $D_s(\rho_{AB}\|\tau_A)$ uses the actual marginal $\rho_B$.
At order one, both quantities coincide with
\begin{equation*}
D_1^*(\rho_{AB}\|\tau_A)
=D_1(\rho_{AB}\|\tau_A)
:=D(\rho_{AB}\|\tau_A\otimes\rho_B).
\end{equation*}
For a classical--quantum state $\rho_{XE}$, the sandwiched and Petz R{\'e}nyi mutual information are respectively defined as
\begin{align*}
I_\alpha^*(X:E)_{\rho_{XE}}
&:=D_\alpha^*(\rho_{XE}\|\rho_X)
=\inf_{\eta_E\in\mathcal S(E)}
D_\alpha^*(\rho_{XE}\|\rho_X\otimes\eta_E),
\\
I_s(X:E)_{\rho_{XE}}
&:=D_s(\rho_{XE}\|\rho_X)
=D_s(\rho_{XE}\|\rho_X\otimes\rho_E).
\end{align*}
Their common order-one value is mutual information
\begin{equation*}
I(X:E)_{\rho_{XE}}:=D(\rho_{XE}\|\rho_X\otimes\rho_E).
\end{equation*}
The sandwiched and Petz R{\'e}nyi conditional entropies are respectively defined as
\begin{equation}\label{eq:entropies}
\begin{aligned}
H_\alpha^*(A|B)_{\rho_{AB}}
&:=-\inf_{\eta_B\in\mathcal S(B)}
D_\alpha^*(\rho_{AB}\|\id_A\otimes\eta_B),
&&\alpha>1,\\
H_s(A|B)_{\rho_{AB}}
&:=-D_s(\rho_{AB}\|\id_A\otimes\rho_B),
&&0\leq s<1.
\end{aligned}
\end{equation}
Both converge at order one to the conditional entropy
\[
H(A|B)_{\rho_{AB}}:=-D(\rho_{AB}\|\id_A\otimes\rho_B).
\]

We record the properties of R{\'e}nyi information quantities needed in the proofs.
\begin{proposition}
Let $\alpha>1$ and $0<s<1$.
The R{\'e}nyi divergences and associated information quantities
satisfy the following properties.
\begin{enumerate}

\item Scaling and tensor products~\cite{hayashi2016correlation,MullerLennert2013}.
Let $\rho\in\mathcal S(A)$ be a state and
$\sigma\in\mathcal P(A)$ a positive semidefinite operator.
For every $c>0$, we have
\begin{align*}
D_\alpha^*(\rho\|c\sigma)=D_\alpha^*(\rho\|\sigma)-\log c,\qquad
D_s(\rho\|c\sigma)
=D_s(\rho\|\sigma)-\log c.
\end{align*}
Both divergences are additive on tensor products.
In particular, for every integer $n\geq1$,
\begin{align*}
D_\alpha^*(\rho^{\otimes n}\|\sigma^{\otimes n})=nD_\alpha^*(\rho\|\sigma),\qquad
D_s(\rho^{\otimes n}\|\sigma^{\otimes n})=nD_s(\rho\|\sigma).
\end{align*}

\item Data processing and order monotonicity~\cite{Beigi2013,FrankLieb2013,MullerLennert2013}.
Let $\rho\in\mathcal S(A)$ and $\sigma\in\mathcal P(A)$.
For every quantum channel $\Phi$, the data-processing inequality gives
\begin{equation*}
D_\alpha^*(\Phi(\rho)\|\Phi(\sigma))
\leq D_\alpha^*(\rho\|\sigma).
\end{equation*}
Moreover, the function $\alpha\mapsto D_\alpha^*(\rho\|\sigma)$
is nondecreasing, and
\begin{equation*}
\lim_{\alpha\to 1}D_\alpha^*(\rho\|\sigma)
=D(\rho\|\sigma).
\end{equation*}

\item Additivity
\cite{hayashi2016correlation}.
Let $\rho_{AB}\in\mathcal S(AB)$ and
$\tau_A\in\mathcal S(A)$ satisfy
$\supp\rho_A\subseteq\supp\tau_A$.
For every integer $n\geq1$, we have
\begin{equation*}
D_\alpha^*(\rho_{AB}^{\otimes n}\|\tau_A^{\otimes n})
=nD_\alpha^*(\rho_{AB}\|\tau_A).
\end{equation*}

\item Regularity in the order~\cite{hayashi2016correlation}.
Let $\rho_{AB}\in\mathcal S(AB)$ and
$\tau_A\in\mathcal S(A)$ satisfy
$\supp\rho_A\subseteq\supp\tau_A$.
The function
$\alpha\mapsto D_\alpha^*(\rho_{AB}\|\tau_A)$
is finite and continuously differentiable for $\alpha\geq1$,
with a right derivative at $\alpha=1$.
By additivity, the same holds for the corresponding
$n$-fold information quantity.
The function
\begin{equation*}
\Lambda(\alpha):=(\alpha-1)D_\alpha^*(\rho_{AB}\|\tau_A),
\qquad \alpha>1,
\qquad \Lambda(1):=0,
\end{equation*}
is convex and continuously differentiable on $[1,\infty)$.
Continuity at order one yields
\begin{equation*}
\Lambda'_+(1)
=D_1^*(\rho_{AB}\|\tau_A)
=D(\rho_{AB}\|\tau_A\otimes\rho_B).
\end{equation*}
These statements apply to arbitrary normalized $\tau_A$
satisfying the support inclusion, including both
$\tau_A=\rho_A$ and $\tau_A=\pi_A$.

\end{enumerate}
\end{proposition}

\subsection{Lemmas for the strong converse exponent}
The strong converse proofs compare one minus the trace-distance error
with a binary testing cost. We collect the testing identities and
asymptotic estimate used in these comparisons.
Let $A,B\geq0$ act on the same finite-dimensional Hilbert space. The noncommutative minimum of $A$ and $B$ is defined as
\begin{equation*}
A\wedge B:=\frac{A+B-|A-B|}{2}.
\end{equation*}
We collect the properties of the noncommutative minimum;
see \cite[Sec.~II, Eq.~(5) and Fact~1]{Cheng2023Achievability} and \cite[Lemma~4]{shen2022strong}.
\begin{lemma}[Properties of the noncommutative minimum]
\label{lem:noncommutative-minimum}\label{lem:pgm}

The following properties hold.
\begin{enumerate}
\item Variational representation and basic bounds.
\begin{equation}\label{eq:testing-cost}
\begin{aligned}
\Tr A\wedge B
&=\min_{0\leq T\leq\id}
\{\Tr A(\id-T)+\Tr BT\}=\frac{\Tr A+\Tr B-\|A-B\|_1}{2}.
\end{aligned}
\end{equation}
\item Monotonicity of the trace under the L\"owner order.
If $A\leq A'$ and $B\leq B'$, then
\begin{equation*}
\Tr A\wedge B\leq\Tr A'\wedge B'.
\end{equation*}

\item Monotone increase under positive trace-preserving
maps.
For every positive trace-preserving linear map $\Phi$, it holds that
\begin{equation*}
\Tr A\wedge B\leq\Tr\Phi(A)\wedge\Phi(B).
\end{equation*}

\item Joint concavity. The map $(A, B)\mapsto \Tr A\wedge B$ is jointly concave.

\item Joint positive homogeneity.
For every $c\geq0$,
\begin{equation*}
\begin{aligned}
(cA)\wedge(cB)&=c(A\wedge B).
\end{aligned}
\end{equation*}

\item Direct sum. If $A,A'$ are positive semidefinite operators on the
same space and $B,B'$ are positive semidefinite operators on a second
space, then
\begin{equation*}
(A\oplus B)\wedge(A'\oplus B')=(A\wedge A')\oplus(B\wedge B').
\end{equation*}
\item Pretty-good test bound.
\begin{equation*}
\Tr\!\left[A(A+B)^{-1/2}B(A+B)^{-1/2}\right]
\leq\Tr A\wedge B.
\end{equation*}
\end{enumerate}
\end{lemma}

The next lemma turns a second-moment estimate into a comparison with
the weighted testing cost $\Tr\rho\wedge m\sigma$.
\begin{lemma}[Testing comparison]\label{lem:test-transfer}
Let $\rho$ be a state, let $\sigma\geq0$, and let $m>0$ and $K\geq0$.
Suppose $0\leq\Delta\leq1$ and a self-adjoint operator $S$ satisfy
\begin{equation*}
-\id\leq S\leq\id, \qquad \Tr\rho S=\Delta,\qquad
 m\Tr\sigma S^2\leq K(1-\Delta).
\end{equation*}
Then
\begin{equation*}
\Tr\rho\wedge m\sigma\leq(2+K)(1-\Delta).
\end{equation*}
\end{lemma}
\begin{proof}
Using the Cauchy--Schwarz inequality, we have
\begin{align*}
|\operatorname{Tr}\rho S|^2=|\operatorname{Tr}\sqrt{\rho}\,S\sqrt{\rho}|^2\le \operatorname{Tr}\rho\,
\operatorname{Tr}\sqrt{\rho}\,S^2\sqrt{\rho}=\operatorname{Tr}\rho S^2.
\end{align*}
Then, $\Tr\rho S^2\geq(\Tr\rho S)^2=\Delta^2$. Consequently,
\begin{equation*}
 \Tr\rho(\id-S^2)\leq1-\Delta^2\leq2(1-\Delta).
\end{equation*}
Therefore, we obtain
\begin{align*}
\Tr \rho\wedge m\sigma\leq \Tr \rho(\id-S^2)+m\Tr \sigma S^2
\leq (2+K)(1-\Delta). & \qedhere
\end{align*}
\end{proof}
For states $\rho,\sigma$ with $\supp\rho\subseteq\supp\sigma$ and any
$R\in\mathbb R$, define
\begin{equation*}
C_R(\rho\|\sigma)
:=\max_{0\le s\le1}(1-s)\{D_s(\rho\|\sigma)-R\}.
\end{equation*}
The next lemma identifies $C_R$ as the exponent of a weighted binary
testing cost. It follows from Hiai, Mosonyi, and Ogawa
\cite[Theorem~3.1 and Corollary~3.3]{HiaiMosonyiOgawa}.

\begin{lemma}[Weighted testing exponent]\label{lem:weighted-testing}
Let $\rho,\sigma\in \mathcal{S}(\mathcal H)$ satisfying
$\supp\rho\subseteq\supp\sigma$. Let \(\{m_n\}_{n\ge1}\) be a sequence of positive real numbers such that
\begin{align*}
\frac{1}{n}\log m_n\to R\quad\text{as }n\to\infty
\end{align*}
for some \(R\in\mathbb R\). Then
\begin{equation}\label{eq:testing-exponent}
\lim_{n\to\infty}-\frac1n\log
\Tr \rho^{\otimes n}\wedge m_n\sigma^{\otimes n}=C_R(\rho\|\sigma).
\end{equation}
\end{lemma}

\begin{proof}
For every $a\in\mathbb R$, Hiai,
Mosonyi, and Ogawa~\cite[Corollary~3.3 and Eq.~(20)]{HiaiMosonyiOgawa}
proved that
\begin{equation}\label{eq:hmo-weighted-exponent}
\lim_{n\to\infty}-\frac1n\log
\Tr e^{-na}\rho^{\otimes n}\wedge \sigma^{\otimes n}
=\max_{0\le s\le1}
\{as-\log\Tr\rho^s\sigma^{1-s}\}.
\end{equation}
By the joint positive homogeneity of
$(A,B)\mapsto\Tr A\wedge B$, we have
\begin{equation*}
\Tr e^{-nR}\rho^{\otimes n}\wedge \sigma^{\otimes n}
=e^{-nR}\Tr \rho^{\otimes n}\wedge e^{nR}\sigma^{\otimes n}.
\end{equation*}
Taking $a=R$ in \eqref{eq:hmo-weighted-exponent} therefore yields
\begin{align}
\lim_{n\to\infty}-\frac1n\log
\Tr \rho^{\otimes n}\wedge e^{nR}\sigma^{\otimes n}
&=\max_{0\le s\le1}
\{-\log\Tr\rho^s\sigma^{1-s}-(1-s)R\}\notag\\
&=\max_{0\le s\le1}(1-s)\{D_s(\rho\|\sigma)-R\}.
\label{eq:fixed-weight-testing-exponent}
\end{align}
To handle the sequence $m_n$, put
\begin{equation*}
\varepsilon_n:=\left|\frac1n\log m_n-R\right|\longrightarrow0.
\end{equation*}
Then $e^{-n\varepsilon_n}e^{nR}\le m_n
\le e^{n\varepsilon_n}e^{nR}$. Since both terms in the variational
definition \eqref{eq:testing-cost} are nonnegative,
\begin{equation*}
e^{-n\varepsilon_n}\Tr \rho^{\otimes n}\wedge e^{nR}\sigma^{\otimes n}
\le\Tr \rho^{\otimes n}\wedge m_n\sigma^{\otimes n}
\le e^{n\varepsilon_n}\Tr \rho^{\otimes n}\wedge e^{nR}\sigma^{\otimes n}.
\end{equation*}
Taking logarithms and dividing by $n$ shows that the normalized
negative logarithms for weights $m_n$ and $e^{nR}$ differ by at most
$\varepsilon_n$, which proves
\eqref{eq:testing-exponent} via
\eqref{eq:fixed-weight-testing-exponent}.
\end{proof}

\subsection{Rosenthal inequalities}
We will use the noncommutative Rosenthal
inequality of Junge and Xu~\cite[Theorem~2.1(i)]{JungeXu2008RosenthalII} for moment bounds for sums of operators. The following special form was derived
in Li, Qiu, and Zhang~\cite{LiQiuZhang2026Reliability}, where it was applied
to reliability functions for quantum soft covering and privacy
amplification under R\'enyi divergence. A reverse inequality for $1<p<2$  was subsequently used 
to study strong converse exponents for the same
tasks~\cite{LiQiuZhang2026StrongConverse}.

\begin{lemma}[Matrix Rosenthal inequality]
\label{lem:rosenthal-moments}
For every $p\geq2$, there is a constant $K_p>0$, depending only on $p$,
such that, for every finite family of independent, mean-zero,
self-adjoint random matrices $Y_i$ with finite $p$-moments, we have
\begin{align*}
\left(
\E\Tr\left|\sum_iY_i\right|^p
\right)^{1/p}
\leq K_p\max\left\{
\left(\sum_i\E\Tr|Y_i|^p\right)^{1/p},
\left[
\Tr\left(\sum_i\E Y_i^2\right)^{p/2}
\right]^{1/p}
\right\}.
\end{align*}
\end{lemma}

The application to convex splitting then requires a weighted conditional version. For simplicity, we first state  Junge and Xu's noncommutative Rosenthal inequality
inequality~\cite[Theorem~2.1(i)]{JungeXu2008RosenthalII} in the matrix setting and deduce the form we need. Let $\mathcal M=\mathcal B(\mathcal{H})$ be a matrix algebra and $\sigma\in \mathcal{S}(\mathcal{H})$ be a faithful state. A $*$-subalgebra $\mathcal N\subseteq\mathcal M$ is called $\sigma$-invariant if
\[
\sigma^{-it}\mathcal N\sigma^{it}=\mathcal N,
\qquad t\in\mathbb R,
\]
Its $\sigma$-preserving conditional expectation
$\mathcal E:\mathcal M\to\mathcal N$ is uniquely characterized by
\[
\Tr (\sigma \mathcal E(a)b)=\Tr (\sigma ab),
\qquad a\in\mathcal M,\quad b\in\mathcal N.
\]
Recall the Kosaki's weighted $L_p$-norm
\[
\|x\|_{p,\sigma}=(\Tr|\sigma^{\frac{1}{2p}} x\sigma^{\frac{1}{2p}}|^p)^{1/p},
\]
and write $L_p(\mathcal M,\sigma)$ for its $L_p$-space. Under the canonical inclusion
$L_p(\mathcal N,\sigma)\subseteq L_p(\mathcal M,\sigma)$,
$\mathcal E$ extends to a contractive projection
\[
\mathcal E_p:L_p(\mathcal M,\sigma)\longrightarrow L_p(\mathcal N,\sigma),
\qquad 1\leq p<\infty.
\]
We omit the subscript $p$ when clear. In this representation,
\[
\mathcal E_p\!\left(\sigma^{1/(2p)}a\sigma^{1/(2p)}\right)
=
\sigma^{1/(2p)}\mathcal E(a)\sigma^{1/(2p)},\quad a\in \mathcal{B}(\mathcal{H}).
\]
We denote the pre-adjoint of
$\mathcal E$ by $\mathcal E_*$. 

A family $\{M_i\}_i$ of $\sigma$-invariant subalgebras
$\mathcal N\subseteq\mathcal M_i\subseteq\mathcal M$
is called independent with respect to
$\mathcal E$ if
\[
\mathcal E(ab)=\mathcal E(a)\mathcal E(b)
\]
for every $i$, every $a\in\mathcal M_i$, and every $b$
in the von Neumann algebra generated by
$\{\mathcal M_j:j\ne i\}$.
A sequence $(x_i)\subset \mathcal M$ is independent
with respect to $\mathcal E$ if
$x_i\in \mathcal M_i$ for such a family.

\begin{lemma}[Noncommutative Rosenthal inequality]
\label{lem:weighted-matrix-rosenthal}
Let $\mathcal M=\mathcal B(\mathcal H)$ be finite dimensional, let
$\sigma$ be a full-rank density matrix, and let
$\mathcal E:\mathcal M\to\mathcal N$ be the $\sigma$-preserving
conditional expectation onto a subalgebra $\mathcal N$.
Fix $2<p<\infty$. Let $(x_i)$ be a finite sequence independent
with respect to $\mathcal E$ such that $\mathcal E(x_i)=0$ for every
$i$, and set $X_i=\sigma^{1/(2p)}x_i\sigma^{1/(2p)}$.
Then there is a constant $K_p\geq1$, depending only on $p$, such that
\begin{equation}
\label{eq:weighted-matrix-rosenthal}
\begin{aligned}
\left\|\sum_i x_i\right\|_{p,\sigma}
\leq K_p\max\Biggl\{
 \left(\sum_i\|x_i\|_{p,\sigma}^p\right)^{1/p},
 \left\|\sum_i\mathcal E_{p/2}(X_i^*X_i)\right\|_{p/2}^{1/2},
 \left\|\sum_i\mathcal E_{p/2}(X_iX_i^*)\right\|_{p/2}^{1/2}
\Biggr\}.
\end{aligned}
\end{equation}
\end{lemma}

The above is \cite[Theorem~2.1(i)]{JungeXu2008RosenthalII}
applied to the Schatten representatives $X_i$. In particular,
$\mathcal E_p(X_i)=0$ follows from $\mathcal E(x_i)=0$, and
$\|X_i\|_p=\|x_i\|_{p,\sigma}$.

\subsection{Lemmas for the reliability function}
The starting point of our reliability
proofs is the conditional hypothesis testing. Recall that
for a positive operator $\sigma$ with distinct spectral projections
$P_1,\ldots,P_v$, its pinching map 
$$\mathcal P_\sigma(X)=\sum_{j=1}^vP_jXP_j$$
is a quantum channel that satisfies the order inequality
$X\leq v\mathcal P_\sigma(X)$ for $X\ge 0$.

We also recall the following properties for permutation-invariant state.
\begin{lemma}[{\cite[Lemma~1]{hayashi2016correlation}}]
\label{lem:universal-state}
Let $| H|=d$. For every $n\in \mathbb{N}$, there exists a permutation-invariant state
$\eta_n\in \mathcal{S}(H^{\otimes n})$ such that
\begin{enumerate}
  \item every permutation-invariant state $\omega\in \mathcal{S}(H^{\otimes n})$ satisfies
  $$ \omega\le\kappa_n\eta_n \qquad \text{with}\qquad \kappa_n=\binom{n+d^2-1}{n}\le(n+1)^{d^2-1}.$$
  \item $\eta_n$ has at most $v_n$ distinct eigenvalues with $v_n=(n+1)^{d-1}$.
  \item $\eta_n$ commutes with every permutation-invariant state, and satisfies
\begin{align*}
U^{\otimes n}\eta_n(U^*)^{\otimes n}=\eta_n
\end{align*}
for every unitary $U\in \mathcal{U}( H)$.
\end{enumerate}
\end{lemma}

Let $\rho_{AB}\in\mathcal S(AB)$ and $\tau_A\in\mathcal S(A)$
satisfy $\supp\rho_A\subseteq\supp\tau_A$.
For any $r\geq0$, define
\begin{equation*}
\mathsf H_r^*(\rho_{AB}\|\tau_A)
:=
\sup_{\alpha>1}\frac{\alpha-1}{\alpha}
\left(r-D_\alpha^*(\rho_{AB}\|\tau_A)\right).
\end{equation*}

 The following conditional testing exponent below follows from the pinching limit of Hayashi and
Tomamichel~\cite[Propositions~8 and~11]{hayashi2016correlation} and the
testing exponent of Mosonyi and Ogawa~\cite[Theorem~IV.5]{MosonyiOgawa2015}.
\begin{theorem}[Conditional testing exponent]
\label{thm:best-testing-exponent}
Let $\rho_{AB}\in\mathcal S(AB)$ and
$\tau_A\in\mathcal S(A)$ satisfy
$\supp\rho_A\subseteq\supp\tau_A$.
For every $r>0$ and integer $n\geq1$, define
\begin{align}
\label{eq:conditional-testing-success}
A_n(r):=&
\sup_{\substack{
  0\leq T\leq\id_{A^nB^n}\\
  \Tr_{A^n}[(\tau_A^{\otimes n}\otimes\id_{B^n})T]
  \leq e^{-nr}\id_{B^n}
}}
\Tr(\rho_{AB}^{\otimes n}T)\\ =&\sup\ \{\ \Tr(\rho_{AB}^{\otimes n}T)\ ;\ 0\leq T\leq\id,\  \Tr[(\tau_A^{\otimes n}\otimes\omega_{B^n})T]
  \leq e^{-nr}\  \forall\   \omega_{B^n} \in \mathcal{S}(B^n)\}
\end{align}
Then, for every $r\geq0$, the limit exists and satisfies
\begin{equation}
\label{eq:conditional-testing-exponent}
\lim_{n\to\infty}-\frac1n\log A_n(r)
=
\mathsf H_r^*(\rho_{AB}\|\tau_A).
\end{equation}
\end{theorem}
\begin{proof}
 We first prove an upper bound on $A_n(r)$.
For every feasible test $T$, every $\alpha>1$, and every
state $\omega_B$, data processing under the binary measurement
$\{T,\id-T\}$, together with tensor additivity, gives
\begin{align*}
\Tr(\rho_{AB}^{\otimes n}T)
\leq
\exp\left[
-\frac{n(\alpha-1)}{\alpha}
\left(
r-D_\alpha^*(\rho_{AB}\|\tau_A\otimes\omega_B)
\right)
\right].
\end{align*}
Optimizing over $\omega_B$ and then over $\alpha>1$ yields
\begin{align*}
A_n(r)\leq e^{-n\mathsf H_r^*(\rho_{AB}\|\tau_A)}.
\end{align*}
Consequently,
\begin{equation}
\label{eq:conditional-testing-converse}
\liminf_{n\to\infty}-\frac1n\log A_n(r)\geq \mathsf H_r^*(\rho_{AB}\|\tau_A).
\end{equation}

For the opposite bound, let $\eta_n$ be the universal state
on $B^n$ from Lemma~\ref{lem:universal-state}, and set
\begin{align*}
\sigma_n=\tau_A^{\otimes n}\otimes\eta_n.
\end{align*}
By \cite[Proposition~8]{hayashi2016correlation}, for every
$\alpha>1$,
\begin{align*}
\lim_{n\to\infty}\frac1n
\log\Tr\left[
\bigl(\mathcal P_{\sigma_n}
(\rho_{AB}^{\otimes n})\bigr)^\alpha
\sigma_n^{1-\alpha}
\right]
=
(\alpha-1)D_\alpha^*(\rho_{AB}\|\tau_A).
\end{align*}
The right-hand side is finite and differentiable in $\alpha>1$
by \cite[Proposition~11]{hayashi2016correlation}.
The achievability part of
\cite[Theorem~IV.5]{MosonyiOgawa2015},
applied to the state sequences
$\{\rho_{AB}^{\otimes n}\}$ and $\{\sigma_n\}$,
therefore implies the following:
for every $\varepsilon>0$, there exist tests
$0\leq S_n\leq\id$ such that
\begin{align*}
\limsup_{n\to\infty}\frac1n\log\Tr(\sigma_nS_n)
&\leq-r-2\varepsilon,\\
\limsup_{n\to\infty}
-\frac1n\log\Tr(\rho_{AB}^{\otimes n}S_n)
&\leq \mathsf H_{r+2\varepsilon}^*(\rho_{AB}\|\tau_A)+\varepsilon.
\end{align*}
In particular, for all sufficiently large $n$,
\begin{align*}
\Tr(\sigma_nS_n)\leq e^{-n(r+\varepsilon)}.
\end{align*}

Averaging $S_n$ over simultaneous permutations of $A^n$
and $B^n$ preserves both trace expressions above.
We may therefore assume that $S_n$ is invariant under
these permutations.
For any state $\omega_{B^n}$, this invariance allows us
to replace $\omega_{B^n}$ by its permutation average in
\begin{align*}
\Tr\bigl[
(\tau_A^{\otimes n}\otimes\omega_{B^n})S_n
\bigr].
\end{align*}
Lemma~\ref{lem:universal-state} bounds that average by
$\kappa_n\eta_n$, where $\log\kappa_n=o(n)$.
Hence, for all sufficiently large $n$,
\begin{align*}
\left\|
\Tr_{A^n}
\bigl[(\tau_A^{\otimes n}\otimes\id_{B^n})S_n\bigr]
\right\|_\infty
=&
\max_{\omega_{B^n}\in\mathcal S(B^n)}
\Tr\bigl[
(\tau_A^{\otimes n}\otimes\omega_{B^n})S_n
\bigr]\notag\\
\leq&
\kappa_n\Tr(\sigma_nS_n)\notag\\
\leq&
\kappa_n e^{-n(r+\varepsilon)}
\leq e^{-nr}.
\end{align*}
Thus $S_n$ is feasible in the definition of $A_n(r)$,
and consequently
\begin{align*}
\limsup_{n\to\infty}-\frac1n\log A_n(r)
\leq \mathsf H_{r+2\varepsilon}^*(\rho_{AB}\|\tau_A)+\varepsilon.
\end{align*}
For fixed $(\rho_{AB},\tau_A)$, the function
$r\mapsto \mathsf H_r^*(\rho_{AB}\|\tau_A)$ is finite and
$1$-Lipschitz, since it is
the supremum of affine functions whose slopes belong
to $(0,1)$.
Letting $\varepsilon\to 0$ and combining the result
with \eqref{eq:conditional-testing-converse} proves
\eqref{eq:conditional-testing-exponent}.
\end{proof}

In the reliability proofs, the testing rate controls the decay of the
acceptance probability and the size of the variance term in the
Rosenthal estimate. The next proposition optimizes the sum of their
contributions to the exponent.

\begin{proposition}
\label{prop:scalar-variational-identity}
Under the assumptions of Theorem~\ref{thm:best-testing-exponent}, for
every $\ell\in\mathbb R$,
\begin{equation}
\label{eq:scalar-testing-identity}
\inf_{r>0}\left\{
\mathsf H_r^*(\rho_{AB}\|\tau_A)+\tfrac12[\ell-r]_+
\right\}
=
\max_{1\leq\alpha\leq2}
\frac{\alpha-1}{\alpha}
\left(\ell-D_\alpha^*(\rho_{AB}\|\tau_A)\right).
\end{equation}
\end{proposition}

\begin{proof}
Let $\Lambda(\alpha)=(\alpha-1)D_\alpha^*(\rho_{AB}\|\tau_A)$ for
$\alpha>1$, with $\Lambda(1)=0$, and define
$F(\alpha)=((\alpha-1)\ell-\Lambda(\alpha))/\alpha$ on $[1,2]$.
For $1<\alpha\leq2$, the defining supremum of $\mathsf H_r^*$ gives
\[
\mathsf H_r^*(\rho_{AB}\|\tau_A)+\tfrac12[\ell-r]_+
\geq
\frac{(\alpha-1)r-\Lambda(\alpha)}{\alpha}
+\frac{\alpha-1}{\alpha}[\ell-r]_+
\geq F(\alpha).
\]
The inequality extends to $\alpha=1$ by continuity.

For the reverse bound, let $\alpha_*$ maximize $F$ on $[1,2]$ and set
$r_*:=\alpha_*\Lambda'(\alpha_*)-\Lambda(\alpha_*)$, using the right
derivative at one. Convexity and $\Lambda(1)=0$ give
$r_*\geq\Lambda'(\alpha_*)\geq D_1^*(\rho_{AB}\|\tau_A)\geq0$.
The tangent inequality for $\Lambda$ at $\alpha_*$ gives, for every
$\beta>1$,
\[
\frac{(\beta-1)r_*-\Lambda(\beta)}{\beta}
\leq r_* - \Lambda'(\alpha_*).
\]
Equality holds at $\beta=\alpha_*$ if $\alpha_*>1$ and in the limit
$\beta\downarrow1$ if $\alpha_*=1$. Hence the right-hand side equals
$\mathsf H_{r_*}^*(\rho_{AB}\|\tau_A)$.

The derivative of $F$ at $\alpha_*$ is $(\ell-r_*)/\alpha_*^2$, with
the appropriate one-sided derivative at an endpoint.
If $1<\alpha_*<2$, then $r_*=\ell$ and the penalty vanishes.
If $\alpha_*=1$, then $\ell\leq r_*$ and both terms vanish.
If $\alpha_*=2$, then $\ell\geq r_*$ and the penalty is
$(\ell-r_*)/2$. In each case
\[
\mathsf H_{r_*}^*(\rho_{AB}\|\tau_A)+\tfrac12[\ell-r_*]_+
=F(\alpha_*).
\]
When $r_*=0$, the same value is approached as $r\downarrow0$
by continuity of $\mathsf H_r^*$ in $r$.
\end{proof}

We next give a general trace-distance lower bound to combine with the
Rosenthal estimates. 

\begin{lemma}[Trace-distance lower bound]\label{general-one-shot}
Let $\mathcal M=\mathcal B(\mathcal H)$, and let
 $\sigma$ be a faithful state. Let
$\mathcal E:\mathcal M\to\mathcal N$
be a $\sigma$-preserving conditional expectation onto the subalgebra $\mathcal N\subset \mathcal M$. Suppose that, for some $L>0$,
\[
-L\sigma\leq\rho-\mathcal E_*(\rho)\leq L\sigma.
\]
Then, for every self-adjoint $S\in\mathcal M$ and $1<p<\infty$,
\[
\left|\Tr[\rho(S-\mathcal E(S))]\right|
\leq L^{1/p}\|\rho-\mathcal E_*(\rho)\|_1^{1-1/p}
\|S-\mathcal E(S)\|_{p,\sigma}.
\]
\end{lemma}
\begin{proof}
Let $\overset{o}{\rho}=\rho-\mathcal E_*(\rho)$ and $q=p/(p-1)$.
Using $\Tr[\mathcal E_*(\rho)X]=\Tr[\rho\mathcal E(X)]$
and $\mathcal E^2=\mathcal E$, we obtain
\[
\Tr[\mathcal E_*(\rho)(S-\mathcal E(S))]
=
\Tr[\rho(\mathcal E(S)-\mathcal E^2(S))]
=0.
\]
Hence
\[
\Tr[\rho(S-\mathcal E(S))]
=
\Tr[\overset{o}{\rho}(S-\mathcal E(S))].
\]
The boundedness assumption implies that
\[
\overset{o}{\rho}=\sigma^{1/2}a\sigma^{1/2},
\qquad
a=a^*\in\mathcal M,
\qquad
\|a\|_\infty\leq L.
\]
Complex interpolation, applied to
$z\mapsto\sigma^{z/2}a\sigma^{z/2}$, gives
\[
\|a\|_{q,\sigma}
\leq
\|a\|_\infty^{1-1/q}\|a\|_{1,\sigma}^{1/q}
\leq
L^{1/p}\|\overset{o}{\rho}\|_1^{1-1/p}.
\]
By traciality and H\"older's inequality,
\begin{align*}
\left|\Tr[\rho(S-\mathcal E(S))]\right|
&\leq
\|a\|_{q,\sigma}\|S-\mathcal E(S)\|_{p,\sigma}\leq
L^{1/p}\|\overset{o}{\rho}\|_1^{1-1/p}
\|S-\mathcal E(S)\|_{p,\sigma}.\qedhere
\end{align*}
\end{proof}
\section{Classical-quantum soft covering}
\label{sec:soft-covering}
\subsection{Problem statement and main result}
\begin{definition}[Classical-Quantum Soft Covering]
Let $
\rho_{XE}=\sum_{x\in\mathcal X}P_X(x)|x\rangle\langle x|\otimes\rho_E^x$
be a classical--quantum state, where $P_X$ is a probability distribution
on a finite alphabet $\mathcal X$ and each $\rho_E^x,x\in \mathcal X$ is a density
operator on a finite-dimensional system $E$. 
Its $E$ marginal is
$\rho_E=\sum_{x\in\mathcal X}P_X(x)\rho_E^x$. For an integer $M\geq1$, a codebook is an indexed tuple
$\mathcal{C}=(x(1),\ldots,x(M))$, with repetitions allowed.
Choosing $m$ uniformly from $\{1,\ldots,M\}$ and sending $x(m)$
through the c-q channel $x\mapsto\rho_E^x$ produces the expected output state
\begin{equation*}
\rho_{E}^{\mathcal C}=\frac{1}{M}\sum_{m=1}^M\rho_{E}^{X(m)}.
\end{equation*}
For soft covering, draw the codewords $X(m)$, $1\leq m\leq M$,
independently from $P_X$. 
Its expected trace-distance error is
$$\Delta=\frac12\mathbb E_\mathcal{C}\left\|\rho_{E}^{\mathcal C}-\rho_E\right\|_1.$$
 \end{definition}
In the $n$-fold setting, the c-q channel is
$x^n\to
\rho_{E^n}^{x^n}:=\bigotimes_{i=1}^n\rho_{E_i}^{x_i}$,
and the target state is $\rho_E^{\otimes n}$. Fix a rate $R\geq0$ and a
sequence of integers $M_n\geq1$ such that $
\displaystyle \lim_{n\to\infty}\frac1n\log M_n=R.$
Draw the codewords of $\mathcal C_n=(X^n(1),\ldots,X^n(M_n))$
independently from $P_X^{\otimes n}$. The expected trace-distance error is
\[
\Delta_n:=\frac12\E_{\mathcal C_n}
\bigl\|\rho_{E^n}^{\mathcal C_n}-\rho_E^{\otimes n}\bigr\|_1.
\]
For i.i.d.\ random codebooks, define the reliability function and
strong converse exponent by
\begin{align*}
E_{\mathtt{sc}}(\rho_{XE},R)&:=\liminf_{n\to\infty}-\frac1n\log\Delta_n,\\
\varGamma_{\mathtt{sc}}(\rho_{XE},R)&:=\liminf_{n\to\infty}-\frac1n\log(1-\Delta_n).
\end{align*}

Exponential estimates for quantum resolvability and its applications
to wiretap coding were studied by Hayashi~\cite{Hayashi2015}.
For classical memoryless channels, Yagli and Cuff determined the exact
exponent of the expected total-variation distance for both i.i.d.\ and
constant-composition random codebooks~\cite{YagliCuff2019}.
In the classical--quantum setting, Cheng and Gao established one-shot
trace-distance bounds yielding achievable error exponents and
exponential strong converse bounds~\cite{ChengGao2024}.
Shen, Gao, and Cheng characterized optimal second-order rates and
moderate-deviation behavior through hypothesis
testing~\cite{ShenGaoCheng2024}, while Hayashi, Cheng, and Gao studied
fixed-input and worst-input resolvability capacities and their strong
converses~\cite{HayashiChengGao2024}.
Related exact exponent results under sandwiched R\'enyi divergence
criteria were obtained by Li, Qiu, and
Zhang~\cite{LiQiuZhang2026Reliability,LiQiuZhang2026StrongConverse}.
Here we study the exact reliability and strong converse exponents
of the expected trace-distance error for i.i.d.\ random codebooks.

\begin{theorem}\label{thm:soft-exponents}
Let $\rho_{XE}=\sum_{x\in\mathcal X}P_X(x)|x\rangle\langle x|\otimes\rho_E^x$
be a c-q state, and let $M_n\geq1$ satisfy
$n^{-1}\log M_n\to R\geq0$. At each blocklength, draw the $M_n$
codewords independently from $P_X^{\otimes n}$. Then the limit exponent
$E_{\mathtt{sc}}$ and $\varGamma_{\mathtt{sc}}$ exist, and
\begin{equation}\label{eq:soft-strong}
\varGamma_{\mathtt{sc}}(\rho_{XE},R)
=\max_{0\leq s\leq1}(1-s) \bigl(I_s(X:E)_\rho-R\bigr).
\end{equation}
\begin{align}\label{eq:soft-reliability}
E_{\mathtt{sc}}(\rho_{XE},R)
=\max_{1\leq\alpha\leq2}\frac{\alpha-1}{\alpha}
 \bigl(R-I_\alpha^*(X:E)_\rho\bigr).
\end{align}
\end{theorem}
For $I(X:E)_\rho>0$, the threshold is $R=I(X:E)_\rho$:
the expected error decays exponentially above this rate and approaches
one exponentially below it.

\begin{remark}[Relation to convex splitting]
\label{rem:soft-convex-splitting}
The random soft-covering error equals the convex-splitting error in
Definition~\ref{def:convex-splitting} with $A=X$, $B=E$, and
$\tau_A=\rho_X$. Indeed, in the block indexed by
$(x(1),\ldots,x(M))$ of the classical registers $X_1,\ldots,X_M$,
the difference of the convex-splitting state and its target is
\[
\left(\prod_{m=1}^M P_X(x(m))\right)
\left(\frac1M\sum_{m=1}^M\rho_E^{x(m)}-\rho_E\right).
\]
The trace norm is additive over these blocks, so
\begin{equation}
\label{eq:soft-convex-identity-n}
\Delta_M(\rho_{XE}\|\rho_X)
=\frac12\E_{\mathcal C}\bigl\|\rho_E^{\mathcal C}-\rho_E\bigr\|_1.
\end{equation}
Here $\Delta_M(\rho_{XE}\|\rho_X)$ is the one-shot convex-splitting
error. At blocklength $n$, the same identity applies to the registers
$X^n,E^n$, the input law $P_X^{\otimes n}$, and $M_n$ codewords.
Since $D_\alpha^*(\rho_{XE}\|\rho_X)=I_\alpha^*(X:E)_\rho$ and
$D_s(\rho_{XE}\|\rho_X)=I_s(X:E)_\rho$, both exponent formulas are
special cases of Theorem~\ref{thm:convex-exponents}.
\end{remark}

\subsection{The strong converse exponent}
\begin{proof}[Proof of \eqref{eq:soft-strong}]
 Using \eqref{eq:soft-convex-identity-n} gives
\begin{equation}
\label{eq:soft-testing}
\frac13\Tr\bigl(\rho_{XE}\wedge M(\rho_X\otimes\rho_E)\bigr)
\leq 1-\frac12\E_{\mathcal C}
\bigl\|\rho_E^{\mathcal C}-\rho_E\bigr\|_1.
\end{equation}
Apply this bound at blocklength $n$ and use
Lemma~\ref{lem:weighted-testing} with weights $M_n$. The result is
\[
\limsup_{n\to\infty}-\frac1n\log(1-\Delta_n)
\leq\max_{0\leq s\leq1}(1-s)\bigl(I_s(X:E)_\rho-R\bigr).
\]
The one-shot bound of Cheng and Gao~\cite{ChengGao2024}
(Theorem~5 of the arXiv version, with $s=2-1/\alpha$) gives
the matching lower bound on the liminf by additivity of the Petz
R\'enyi divergence. The two bounds establish the limit and
\eqref{eq:soft-strong}.
\end{proof}

\subsection{The reliability function}

We give a direct reliability proof because the c-q structure yields a
one-shot estimate that is independent of the smallest positive input
probability.
\begin{lemma}
\label{lem:soft-testing-lower}
Let $M\geq1$, and suppose that $0\leq T_x\leq\id_E$ satisfy
\[
\E_X T_X\leq b\id_E.
\]
Then, for every $2\leq p<\infty$,
\begin{equation}
\label{eq:soft-testing-lower}
\frac12\E_{\mathcal C}
\left\|\rho_E^{\mathcal C}-\rho_E\right\|_1
\geq
\frac12
\left(
\frac{
[\E_X\Tr(\rho_E^XT_X)-b]_+
}{
K_p|E|^{1/p}\max\{1,\sqrt{Mb}\}
}
\right)^{p/(p-1)},
\end{equation}
where $K_p$ is the constant in
Lemma~\ref{lem:rosenthal-moments}.
\end{lemma}

\begin{proof}
We apply Lemma~\ref{general-one-shot} with
\[
\mathcal M
=L_\infty(\mathcal X^M,P_X^{\otimes M})
\overline\otimes\mathcal B(E),\quad \mathcal{N}=\mathcal B(E),
\qquad
\mathcal E=\E_{\mathcal C},
\]
and
\[
S=\sum_{i=1}^M T_{X(i)},\qquad\rho=\rho_E^{\mathcal C},\qquad \varphi=\id_E/|E|.\]
We equip $\mathcal M$ with the trace $\E_{\mathcal C}\Tr$.
The expectation $\E_{\mathcal C}$ preserves $\varphi$ and maps
$\rho_E^{\mathcal C}$ to $\rho_E$.
Since
\[
-\id_E\leq\rho_E^{\mathcal C}-\rho_E\leq\id_E,
\]
the boundedness assumption in Lemma~\ref{general-one-shot}
holds with $L=|E|$. Lemma~\ref{general-one-shot} therefore gives
\begin{align}
\label{sc-one}
\left|\E_{\mathcal C}\Tr[\rho_E^{\mathcal C}(S-\E_{\mathcal C}S)]\right|\leq
\bigl(\E_{\mathcal C}\|\rho_E^{\mathcal C}-\rho_E\|_1\bigr)^{1-1/p}
|E|^{1/p}\|S-\E_{\mathcal C}S\|_{p,\varphi}.
\end{align}

For the left-hand side of \eqref{sc-one}, independence of the codewords gives
\[
\E_{\mathcal C}\Tr\!\left[
\rho_E^{\mathcal C}(S-\E_{\mathcal C}S)
\right]=\E_X\Tr(\rho_E^XT_X)
-\Tr(\rho_E\E_XT_X)\geq\E_X\Tr(\rho_E^XT_X)-b.
\]
To bound the weighted norm, the matrices $T_{X(i)}-\E_XT_X$
are independent, have expectation zero, and have operator
norm at most one. Thus, for $p\geq2$, we have
\begin{align*}
\left(
\sum_{i=1}^M\E_{\mathcal C}
\Tr|T_{X(i)}-\E_XT_X|^p
\right)^{1/p}\leq
\left(
M\Tr\bigl[\E_XT_X^2-(\E_XT_X)^2\bigr]
\right)^{1/p}\leq (M|E|b)^{1/p},
\end{align*}
and
\begin{align*}
\left[
\Tr\left(
\sum_{i=1}^M\E_{\mathcal C}
(T_{X(i)}-\E_XT_X)^2
\right)^{p/2}
\right]^{1/p}=
\sqrt M\left[
\Tr\bigl(\E_XT_X^2-(\E_XT_X)^2\bigr)^{p/2}
\right]^{1/p}\leq |E|^{1/p}\sqrt{Mb}.
\end{align*}
Here we used $0\leq T_x\leq\id_E$ and
$0\leq \E_XT_X^2\leq \E_XT_X\leq b\id_E$.
Applying Lemma~\ref{lem:rosenthal-moments} with $Y_i=T_{X(i)}-\E_XT_X$ yields
\begin{align*}
\|S-\E_{\mathcal C} S\|_{p,\varphi}
&=|E|^{-1/p}
\left(
\E_{\mathcal C}\Tr
\left|\sum_{i=1}^M(T_{X(i)}-\E_XT_X)\right|^p
\right)^{1/p}\\
&\leq K_p\max\{(Mb)^{1/p},\sqrt{Mb}\}\\
&\leq K_p\max\{1,\sqrt{Mb}\}.
\end{align*}
The absolute value in \eqref{sc-one} bounds the positive part
$[\E_X\Tr(\rho_E^XT_X)-b]_+$ from above. Substituting the moment
bound into \eqref{sc-one} and rearranging proves the claim.
\end{proof}

\begin{proof}[Proof of \eqref{eq:soft-reliability}]
Assume $I(X:E)_\rho>0$.
The soft-covering estimate of Cheng and
Gao~\cite{ChengGao2024} (Theorem~1 of the arXiv version), together
with tensor additivity
of $I_\alpha^*(X:E)_\rho$, gives
\[
\Delta_n
\leq
\exp\left[
-\frac{\alpha-1}{\alpha}
\left(\log M_n-nI_\alpha^*(X:E)_\rho\right)
\right],
\qquad 1<\alpha<2.
\]
Taking negative logarithms, dividing by $n$, and
optimizing over $\alpha$ yields
\begin{equation}
\label{eq:soft-matrix-direct-exponent}
\liminf_{n\to\infty}
-\frac1n\log
\Delta_n
\geq
\max_{1\leq\alpha\leq2}
\frac{\alpha-1}{\alpha}
\left(R-I_\alpha^*(X:E)_\rho\right),
\end{equation}
where the endpoint values are included by continuity.

For the matching upper bound on the reliability exponent, we apply
Lemma~\ref{lem:soft-testing-lower} to an optimal conditional test.
The testing pair in Theorem~\ref{thm:best-testing-exponent} is
$(\rho_{XE},\rho_X)$; write $A_n(r)$ for its optimal success
probability.
In this case,
\[
D_\alpha^*(\rho_{XE}\|\rho_X)
=
I_\alpha^*(X:E)_\rho,
\qquad
\mathsf H_r^*(\rho_{XE}\|\rho_X)
=
\sup_{\alpha>1}
\frac{\alpha-1}{\alpha}
\left(r-I_\alpha^*(X:E)_\rho\right).
\]
Fix $r>0$.
The feasible set in the definition of $A_n(r)$ is compact and
contains $e^{-nr}\id_{X^nE^n}$.
Consequently, the maximum is attained and
\begin{equation*}
A_n(r)\geq e^{-nr}>0.
\end{equation*}
Dephasing a maximizing test in the classical basis of $X^n$
preserves its objective and its constraint. Thus we may take
$T_n=\sum_{x^n}|x^n\rangle\langle x^n|\otimes T_{n,x^n}$, where
$0\leq T_{n,x^n}\leq\id_{E^n}$ and
\[
\sum_{x^n}P_X^{\otimes n}(x^n)T_{n,x^n}\leq e^{-nr}\id_{E^n},
\qquad
\sum_{x^n}P_X^{\otimes n}(x^n)
\Tr(\rho_{E^n}^{x^n}T_{n,x^n})=A_n(r).
\]
For every fixed $p\geq2$,
Lemma~\ref{lem:soft-testing-lower} now gives
\begin{equation}
\label{eq:soft-testing-profile}
\Delta_n
\geq
\frac12
\left(
\frac{
[A_n(r)-e^{-nr}]_+
}{
K_p|E|^{n/p}
\max\{1,\sqrt{M_ne^{-nr}}\}
}
\right)^{p/(p-1)}.
\end{equation}
Monotonicity of the sandwiched R\'enyi information
in its order implies
\begin{equation}\label{eq:s-c-1}
\mathsf H_r^*(\rho_{XE}\|\rho_X)\leq[r-I(X:E)_\rho]_+<r.
\end{equation}
By Theorem~\ref{thm:best-testing-exponent}, we have
\begin{equation}\label{eq:s-c-2}
A_n(r)=\exp[-n(\mathsf H_r^*(\rho_{XE}\|\rho_X)+o(1))].
\end{equation}
Combining \eqref{eq:s-c-1} and \eqref{eq:s-c-2} yields
\[
\lim_{n\to\infty}-\frac1n\log[A_n(r)-e^{-nr}]_+
= \mathsf H_r^*(\rho_{XE}\|\rho_X).
\]
For fixed $p$, we also have
\[
\lim_{n\to\infty}\frac1n\log\max\{1,\sqrt{M_ne^{-nr}}\}
=\frac12[R-r]_+.
\]
Taking negative logarithms in \eqref{eq:soft-testing-profile}, dividing
by $n$, and letting $n\to\infty$ with $p$ fixed gives
\begin{align*}
&\limsup_{n\to\infty}
-\frac1n\log
\Delta_n\leq
\frac{p}{p-1}
\left(
\mathsf H_r^*(\rho_{XE}\|\rho_X)+\frac12[R-r]_+
+\frac{\log|E|}{p}
\right).
\end{align*}
Letting $p\to\infty$
therefore yields
\begin{equation}
\label{eq:soft-testing-exponent-bound}
\limsup_{n\to\infty}
-\frac1n\log
\Delta_n
\leq
\mathsf H_r^*(\rho_{XE}\|\rho_X)+\frac12[R-r]_+,
\qquad r>0.
\end{equation}
Taking the infimum over $r>0$ in
\eqref{eq:soft-testing-exponent-bound} and applying
\eqref{eq:scalar-testing-identity} with $\ell=R$ gives
\[
\limsup_{n\to\infty}-\frac1n\log\Delta_n
\leq\max_{1\leq\alpha\leq2}\frac{\alpha-1}{\alpha}
\left(R-I_\alpha^*(X:E)_\rho\right).
\]
Together with \eqref{eq:soft-matrix-direct-exponent},
this proves that the limit exists and establishes
\eqref{eq:soft-reliability}.
\end{proof}

\section{Convex splitting}
\label{sec:convex-splitting}
\subsection{Problem statement and main result}
\begin{definition}[Convex splitting with uniform randomness]
\label{def:convex-splitting}
Let $\rho_{AB}$ and $\tau_A$ be density operators on finite-dimensional
systems satisfying $\operatorname{supp}\rho_A\subseteq\operatorname{supp}\tau_A$.
Let $M\geq1$ be an integer and write $[M]:=\{1,2,\ldots,M\}$.
Alice and Bob share a bipartite state $\rho_{AB}$, and Alice chooses $m\in[M]$ uniformly
and places her part of $\rho_{AB}$ in register $A_m$, where
$A_1,\ldots,A_M$ are each isomorphic to $A$. She prepares the other
$M-1$ registers independently in $\tau_A$ and discards the index $m$
leaves the averaged state
\begin{equation*}
\omega_{A_1\cdots A_MB}
:=\frac1M\sum_{m=1}^M\rho_{A_mB}\otimes
\tau_A^{\otimes[M]\setminus\{m\}}.
\end{equation*}
The task is to approximate the product state
$\bigotimes_{m=1}^M\tau_{A_m}\otimes\rho_B$ with the smallest $M$ for a
given error tolerance. The trace-distance error is
\begin{equation*}
\Delta_M(\rho_{AB}\|\tau_A)
:=\frac12\left\|\omega_{A_1\cdots A_MB}
-\tau_A^{\otimes M}\otimes\rho_B\right\|_1.
\end{equation*}
\end{definition}

Convex splitting was introduced by Anshu, Devabathini, and
Jain~\cite{AnshuDevabathiniJain2017} in their coherent rejection
sampling approach to quantum communication.
Anshu, Jain, and Warsi used convex splitting together with
hypothesis testing and position-based decoding for one-shot
measurement compression and randomness
extraction~\cite{AnshuJainWarsi2019Measurement}.
Cheng and Gao established one-shot error and strong converse bounds
in trace distance, together with matched second-order asymptotics
for its sample complexity~\cite{ChengGao2023Convex}.
Berta, Cheng, and Gao extended the method to multipartite convex
splitting, obtaining trace-distance error exponent bounds and
applications to quantum broadcast channel
simulation~\cite{BertaChengGao2025Multipartite}.
We determine the exact reliability and strong converse exponents of
uniform convex splitting in the i.i.d.\ setting.

In the i.i.d.\ setting, apply Definition~\ref{def:convex-splitting}
to the input $\rho_{AB}^{\otimes n}$ and reference $\tau_A^{\otimes n}$.
Place the entire input subsystem $A^n$ in one of $M_n\geq1$ registers
$A_1^n,\ldots,A_{M_n}^n$, chosen uniformly, with
$n^{-1}\log M_n\to R\geq0$. The splitting state is
\begin{equation*}
\omega_{A_1^n\cdots A_{M_n}^nB^n}^{(n)}
:=\frac1{M_n}\sum_{m=1}^{M_n}
\rho_{A_mB}^{\otimes n}\otimes
\bigotimes_{\substack{1\leq k\leq M_n\\k\ne m}}
\tau_{A_k}^{\otimes n}.
\end{equation*}
The target product state is
\begin{equation*}
\left(\bigotimes_{m=1}^{M_n}\tau_{A_m}^{\otimes n}\right)
\otimes\rho_B^{\otimes n}
=(\tau_A^{\otimes n})^{\otimes M_n}\otimes\rho_B^{\otimes n}.
\end{equation*}
The subscripts in these expressions specify the registers on which
each factor acts. The approximation error is
\begin{equation*}
\Delta_{M_n}(\rho_{AB}^{\otimes n}\|\tau_A^{\otimes n})
:=\frac12\left\|\omega_{A_1^n\cdots A_{M_n}^nB^n}^{(n)}
-(\tau_A^{\otimes n})^{\otimes M_n}\otimes\rho_B^{\otimes n}
\right\|_1.
\end{equation*}
For brevity, write
$\Delta_n:=\Delta_{M_n}(\rho_{AB}^{\otimes n}\|\tau_A^{\otimes n})$.

Define the reliability function and strong converse exponent by
\begin{align*}
E_{\mathtt{cs}}(\rho_{AB}\|\tau_A,R)
&:=\liminf_{n\to\infty}-\frac1n\log\Delta_n,\\
\varGamma_{\mathtt{cs}}(\rho_{AB}\|\tau_A,R)
&:=\liminf_{n\to\infty}-\frac1n\log(1-\Delta_n).
\end{align*}

\begin{theorem}\label{thm:convex-exponents}
Let $\rho_{AB},\tau_A$ be states satisfying
$\supp\rho_A\subseteq\supp\tau_A$, and let $M_n\geq1$ satisfy
$\lim_{n\to\infty }n^{-1}\log M_n= R$.
The limit exponent $E_{\mathtt{cs}}$ and $\varGamma_{\mathtt{cs}}$
exist, and 
\begin{equation}\label{eq:convex-sc}
\varGamma_{\mathtt{cs}}(\rho_{AB}\|\tau_A,R)
= \max_{0\leq s\leq1}(1-s)\left( D_s(\rho_{AB}\|\tau_A)-R\right),
\end{equation}
\begin{equation}\label{eq:convex-reliability}
E_{\mathtt{cs}}(\rho_{AB}\|\tau_A,R)
 =\max_{1\leq\alpha\leq2}\frac{\alpha-1}{\alpha}(R-D_\alpha^*(\rho_{AB}\|\tau_A)).
\end{equation}
\end{theorem}
For $\rho_{AB}\ne\tau_A\otimes\rho_B$, the threshold is
$R=D(\rho_{AB}\|\tau_A\otimes\rho_B)$. The error decays exponentially
when the rate $R$ exceeds this threshold and approaches one
exponentially when $R$ is below it.

\subsection{The strong converse exponent}
The one-shot strong converse bound of Cheng and
Gao~\cite[Theorem~7, Eq.~(116), arXiv version]{ChengGao2023Convex},
applied to the $n$-copy input, gives for every $0\leq s\leq1$
\begin{equation*}
1-\Delta_n\leq4\exp\left[-(1-s)
\left(nD_s(\rho_{AB}\|\tau_A)-\log M_n\right)\right].
\end{equation*}
Consequently,
\begin{equation}\label{eq:convex-sc-direct-limit}
\liminf_{n\to\infty}-\frac1n\log(1-\Delta_n)
\geq\max_{0\leq s\leq1}(1-s)
\bigl(D_s(\rho_{AB}\|\tau_A)-R\bigr).
\end{equation}
We prove the matching upper bound.
\begin{proof}[Proof of \eqref{eq:convex-sc}]
By Lemma~\ref{lem:weighted-testing}, the desired upper bound on the
strong converse exponent follows once we prove, for every integer
$M\geq1$,
\begin{equation}\label{eq:convex-testing-reverse}
\frac13\Tr \rho_{AB}\wedge M\tau_A\otimes\rho_B
\leq1-\Delta_M(\rho_{AB}\|\tau_A).
\end{equation}
We prove this comparison using
Lemma~\ref{lem:test-transfer} with $\rho=\rho_{AB}$,
$\sigma=\tau_A\otimes\rho_B$, $m=M$, $K=1$, and
$\Delta=\frac12\|\omega_{A_1\cdots A_MB}-\tau_A^{\otimes M}\otimes\rho_B\|_1$.
We use the weighted partial trace $\mathcal E_\tau$ and its
tensor-product extension,
\begin{align*}
\mathcal E_\tau(X)=\Tr_A[(\tau_A\otimes\id_B)X],\qquad
\mathcal E_{\tau^{\otimes M}}(X)=\Tr_{A_1\cdots A_M}
[(\tau_A^{\otimes M}\otimes\id_B)X].
\end{align*}
Let $\Pi=\{\omega_{A_1\cdots A_MB}-\tau_A^{\otimes M}\otimes\rho_B>0\}$ be the spectral projection
onto the strictly positive spectral subspace.
Identifying $A_1$ with $A$, define
\begin{equation*}
S=\Tr_{A_2\cdots A_M}\left[
(\id_{A_1}\otimes\tau_A^{\otimes(M-1)}\otimes\id_B)\Pi
\right]
-\id_A\otimes\mathcal E_{\tau^{\otimes M}}(\Pi).
\end{equation*}
The partial trace over an empty collection of registers is
understood as the identity map. By construction,
$\mathcal E_\tau(S)=0$.
Since $\Pi$ is invariant under permutations of the $A$ registers,
its weighted partial trace retaining any $A_jB$, after identifying
$A_j$ with $A$, equals
\begin{equation*}
S+\id_A\otimes\mathcal E_{\tau^{\otimes M}}(\Pi).
\end{equation*}
Since
\begin{align*}
0\leq \Tr_{A_2\cdots A_M}\left[
(\id_{A_1}\otimes\tau_A^{\otimes(M-1)}\otimes\id_B)\Pi
\right] \leq \id_{AB}, \qquad 0\leq \mathcal E_{\tau^{\otimes M}}(\Pi) \leq \id_{B},
\end{align*}
$S$ is self-adjoint and
$-\id\leq S\leq\id$. Furthermore, permutation invariance and the
definition of $\Pi$ give
\begin{align}
\Tr\rho_{AB}S
&=\Tr\omega_{A_1\cdots A_MB}\Pi
-\Tr\rho_B\mathcal E_{\tau^{\otimes M}}(\Pi)\notag\\
&=\Tr(\omega_{A_1\cdots A_MB}-\tau_A^{\otimes M}\otimes\rho_B)\Pi\notag\\
&=\Delta_M(\rho_{AB}\|\tau_A).
\label{eq:convex:anova-source}
\end{align}

We next prove the operator inequality
\begin{equation}\label{eq:convex:anova}
M\mathcal E_\tau(S^2)
\leq\mathcal E_{\tau^{\otimes M}}(\Pi)
-\bigl(\mathcal E_{\tau^{\otimes M}}(\Pi)\bigr)^2.
\end{equation}
Let $\pi_j(S)$ denote the insertion of $S$ into $A_jB$,
with identity operators on the other $A$ registers.
For $j\ne k$, taking the weighted partial traces over $A_j$ and
$A_k$ first gives
\begin{equation*}
\mathcal E_{\tau^{\otimes M}}[\pi_j(S)\pi_k(S)]
=\mathcal E_\tau(S)\mathcal E_\tau(S)=0.
\end{equation*}
For each $j$, we first take the weighted partial trace over
all $A$ registers except $A_j$, and identify the retained
register $A_j$ with $A$.
Since $\pi_j(S)$ acts only on $A_jB$, it can be taken outside
this partial trace on the right. Thus,
\begin{align*}
&\mathcal E_{\tau^{\otimes M}}\!\left[
\bigl(\Pi-\id_{A^M}\otimes
\mathcal E_{\tau^{\otimes M}}(\Pi)\bigr)\pi_j(S)
\right]\notag\\
=&
\mathcal E_\tau\!\left[
\left(
\Tr_{A_1\cdots A_{j-1}A_{j+1}\cdots A_M}\!\left[
\left(
\bigotimes_{k=1}^{j-1}\tau_{A_k}
\otimes\id_{A_j}
\otimes\bigotimes_{k=j+1}^{M}\tau_{A_k}
\otimes\id_B
\right)\Pi
\right]
-\id_A\otimes\mathcal E_{\tau^{\otimes M}}(\Pi)
\right)S
\right]\notag\\
=&\mathcal E_\tau(S^2).
\end{align*}
The last equality follows from the permutation invariance
of $\Pi$ and the definition of $S$.
Similarly, we obtain
\begin{align*}
\mathcal E_{\tau^{\otimes M}}\left[
\pi_j(S)\bigl(\Pi-\id_{A^M}\otimes
\mathcal E_{\tau^{\otimes M}}(\Pi)\bigr)
\right]
=\mathcal E_\tau(S^2).
\end{align*}
Also,
$\mathcal E_{\tau^{\otimes M}}[\pi_j(S)^2]
=\mathcal E_\tau(S^2)$.
Consequently, using $\Pi^2=\Pi$ and the preceding identities,
we obtain
\begin{align*}
0
&\leq\mathcal E_{\tau^{\otimes M}}\left[
\left(\Pi-\id_{A^M}\otimes
\mathcal E_{\tau^{\otimes M}}(\Pi)
-\sum_{j=1}^M\pi_j(S)\right)^2
\right] \notag\\
&=\mathcal E_{\tau^{\otimes M}}(\Pi)
-\bigl(\mathcal E_{\tau^{\otimes M}}(\Pi)\bigr)^2
-M\mathcal E_\tau(S^2).
\end{align*}
This proves \eqref{eq:convex:anova}. The two nonnegative testing errors
associated with $\Pi$ satisfy
\begin{align*}
1-\Delta_M(\rho_{AB}\|\tau_A)
&=\Tr\omega_{A_1\cdots A_MB}(\id-\Pi)
+\Tr\rho_B\mathcal E_{\tau^{\otimes M}}(\Pi)\\
&\geq\Tr\rho_B\mathcal E_{\tau^{\otimes M}}(\Pi).
\end{align*}
Combining this with \eqref{eq:convex:anova} gives
\begin{align*}
M\Tr[(\tau_A\otimes\rho_B)S^2]
&=M\Tr\rho_B\mathcal E_\tau(S^2)\notag\\
&\leq\Tr\rho_B\left[
\mathcal E_{\tau^{\otimes M}}(\Pi)
-\bigl(\mathcal E_{\tau^{\otimes M}}(\Pi)\bigr)^2
\right]\notag\\
&\leq\Tr\rho_B\mathcal E_{\tau^{\otimes M}}(\Pi)\notag\\
&\leq1-\Delta_M(\rho_{AB}\|\tau_A).
\end{align*}
Together with $-\id\leq S\leq\id$ and
\eqref{eq:convex:anova-source}, this verifies all the hypotheses
of Lemma~\ref{lem:test-transfer} with the parameters specified
above. That lemma yields \eqref{eq:convex-testing-reverse}.

Finally, the support assumption ensures
\begin{equation*}
\supp\rho_{AB}
\subseteq\supp\rho_A\otimes\supp\rho_B
\subseteq\supp\tau_A\otimes\supp\rho_B
=\supp(\tau_A\otimes\rho_B).
\end{equation*}
Applying \eqref{eq:convex-testing-reverse} to the $n$-copy
splitting state gives
\begin{equation*}
\frac13\Tr\rho_{AB}^{\otimes n}\wedge
M_n(\tau_A\otimes\rho_B)^{\otimes n}
\leq1-\Delta_n.
\end{equation*}
The factor $1/3$ contributes only $n^{-1}\log3$ to the normalized
logarithm. Lemma~\ref{lem:weighted-testing} therefore yields
\begin{align*}
\limsup_{n\to\infty}-\frac1n\log(1-\Delta_n)
&\leq\lim_{n\to\infty}-\frac1n\log
\Tr\rho_{AB}^{\otimes n}\wedge
M_n(\tau_A\otimes\rho_B)^{\otimes n}\\
&=\max_{0\leq s\leq1}(1-s)
\bigl(D_s(\rho_{AB}\|\tau_A)-R\bigr).
\end{align*}
This bound and \eqref{eq:convex-sc-direct-limit} sandwich the
liminf and limsup at the right-hand side of~\eqref{eq:convex-sc}.
Thus the limit exists and \eqref{eq:convex-sc} follows.
\end{proof}

\subsection{The reliability function}
We restrict \(A\) to \(\supp\tau_A\) in this subsection. By the
support assumption, this preserves the splitting state and its target
up to isometric embedding. Compressing a test to this support or
extending it by zero also preserves its conditional constraint and
acceptance probability.

We first lower-bound the splitting error in terms of a conditional test.
\begin{lemma}
\label{lem:convex-testing-lower}
For every $p>2$, let $K_p$ be the constant in
Lemma~\ref{lem:weighted-matrix-rosenthal}. For every integer $M\geq1$,
every $0<b\leq1$, and every test $0\leq T\leq\id_{AB}$ satisfying
\[
\Tr_A[(\tau_A\otimes\id_B)T]\leq b\id_B,
\]
we have
\begin{equation}
\label{eq:convex-testing-lower}
\Delta_M(\rho_{AB}\|\tau_A)
\geq\frac12
\left(
\frac{[\Tr(\rho_{AB}T)-b]_+}
{K_p(|B|/\lambda^{3/2})^{1/p}
\max\{M^{1/p},\sqrt{Mb}\}}
\right)^{p/(p-1)}
\end{equation}
where $\lambda$ is the minimal eigenvalue of $\tau_A$.
\end{lemma}

\begin{proof}
We apply Lemma~\ref{general-one-shot} with
\[
\mathcal M=\mathcal B(A_1\cdots A_MB),
\qquad
\mathcal N=\id_{A_1\cdots A_M}\otimes\mathcal B(B),
\]
\[
\mathcal E(X)
=
\id_{A_1\cdots A_M}\otimes
\Tr_{A_1\cdots A_M}
\bigl[(\tau_A^{\otimes M}\otimes\id_B)X\bigr],
\]
and
\[
S=\sum_{i=1}^M T_{A_iB},
\qquad
\rho=\omega_{A_1\cdots A_MB},
\qquad
\varphi=\tau_A^{\otimes M}\otimes \frac{\id_B}{|B|}.
\]
Here operators on $A_iB$ act as the identity on the remaining systems.
The expectation $\mathcal E$ preserves
$\tau_A^{\otimes M}\otimes\frac{\id_B}{|B|}$, and
$\mathcal E_*(\omega_{A_1\cdots A_MB})=\tau_A^{\otimes M}\otimes\rho_B$.
Since
\[
\rho_{AB}\le\id_{AB}
\le\lambda^{-1}\tau_A\otimes\id_B,
\qquad
\rho_B\le\id_B,
\]
both states are bounded above by $(|B|/\lambda)\varphi$.
Thus the boundedness assumption holds with $L=|B|/\lambda$.
The multiplicative form of Lemma~\ref{general-one-shot}, using
$\|\omega_{A_1\cdots A_MB}-\mathcal E_*(\omega_{A_1\cdots A_MB})\|_1
=2\Delta_M(\rho_{AB}\|\tau_A)$, gives
\begin{equation}
\label{cs-one}
\left|\Tr\bigl[
\omega_{A_1\cdots A_MB}(S-\mathcal E(S))
\bigr]\right|
\leq (|B|/\lambda)^{1/p}
\bigl(2\Delta_M(\rho_{AB}\|\tau_A)\bigr)^{1-1/p}
\|S-\mathcal E(S)\|_{p,\varphi}.
\end{equation}

Expanding the splitting state gives
\[
\begin{aligned}
\Tr\bigl[
\omega_{A_1\cdots A_MB}(S-\mathcal E(S))
\bigr]=
\Tr(\rho_{AB}T)
-
\Tr\!\left[
\rho_B\Tr_A\bigl[(\tau_A\otimes\id_B)T\bigr]
\right]\ge \Tr(\rho_{AB}T)-b.
\end{aligned}
\]
To bound the weighted norm, we use the fact that the algebras
$\mathcal B(A_iB)$ are independent with respect to $\mathcal E$.
Since $0\le T_{A_iB},\mathcal E(T_{A_iB})\le\id$, we have
\[
\begin{aligned}
\left(
\sum_{i=1}^M
\left\|
\varphi^{1/(2p)}
\bigl(T_{A_iB}-\mathcal E(T_{A_iB})\bigr)
\varphi^{1/(2p)}
\right\|_p^p
\right)^{1/p}\le
M^{1/p}\|\varphi^{1/p}\|_{p}
=
M^{1/p}.
\end{aligned}
\]
Using $\tau_A^{1/p}\le\id_A$ and
$\tau_A^{1-1/p}\le\lambda^{-1/p}\tau_A$
in the weighted partial trace, we also obtain
\[
\begin{aligned}
&\left\|
\left(
\sum_{i=1}^M
\mathcal E_{p/2}\!\left[
\left(
\varphi^{1/(2p)}
\bigl(T_{A_iB}-\mathcal E(T_{A_iB})\bigr)
\varphi^{1/(2p)}
\right)^2
\right]
\right)^{1/2}
\right\|_p\\
\le&
\lambda^{-1/(2p)}
\left\|
\left(
\varphi^{1/p}
\sum_{i=1}^M
\bigl[
\mathcal E(T_{A_iB}^2)-\mathcal E(T_{A_iB})^2
\bigr]
\varphi^{1/p}
\right)^{1/2}
\right\|_p\\
\le&
\lambda^{-1/(2p)}\sqrt{Mb}\,
\|\varphi^{1/p}\|_p
=
\lambda^{-1/(2p)}\sqrt{Mb},
\end{aligned}
\]
where we used $0\leq T_{A_iB}\leq\id$ and
$0\leq\mathcal E(T_{A_iB}^2)\leq\mathcal E(T_{A_iB})\leq b\id$.
Applying Lemma~\ref{lem:weighted-matrix-rosenthal} to
$\varphi^{1/(2p)}
\bigl(T_{A_iB}-\mathcal E(T_{A_iB})\bigr)
\varphi^{1/(2p)}$ yields
\begin{align*}
\|S-\mathcal E(S)\|_{p,\varphi}=
\left\|
\sum_{i=1}^M
\varphi^{1/(2p)}
\bigl(T_{A_iB}-\mathcal E(T_{A_iB})\bigr)
\varphi^{1/(2p)}
\right\|_p\leq
K_p\max\left\{M^{1/p},\lambda^{-1/(2p)}\sqrt{Mb}\right\}.
\end{align*}
Since $\lambda\le1$, it follows that
\[
(|B|/\lambda)^{1/p}\|S-\mathcal E(S)\|_{p,\varphi}
\le
K_p\left(|B|/\lambda^{3/2}\right)^{1/p}
\max\left\{M^{1/p},\sqrt{Mb}\right\}.
\]
Let
$C=K_p(|B|/\lambda^{3/2})^{1/p}
\max\{M^{1/p},\sqrt{Mb}\}>0$.
The numerator estimate, \eqref{cs-one}, and the Rosenthal bound give
\[
[\Tr(\rho_{AB}T)-b]_+
\leq C\bigl(2\Delta_M(\rho_{AB}\|\tau_A)\bigr)^{1-1/p}.
\]
Rearranging proves \eqref{eq:convex-testing-lower}, including when
$T=0$ or the centered norm vanishes.
\end{proof}

\begin{proof}[Proof of \eqref{eq:convex-reliability}]
We first prove the upper bound on the reliability exponent by applying
Lemma~\ref{lem:convex-testing-lower} to an optimal conditional test.
The matching lower bound follows from the known achievable estimate.
Assume $D_1^*(\rho_{AB}\|\tau_A)
=D(\rho_{AB}\|\tau_A\otimes\rho_B)>0$.
Let $M_n\geq1$ satisfy $n^{-1}\log M_n\to R\geq0$.
Fix $p>2$ and $r>0$. We choose a maximizing test $T_n$ in Theorem~\ref{thm:best-testing-exponent} such that
\[
\Tr(\rho_{AB}^{\otimes n}T_n)=A_n(r),
\qquad
\Tr_{A^n}[(\tau_A^{\otimes n}\otimes\id_{B^n})T_n]
\leq e^{-nr}\id_{B^n}.
\]
The smallest eigenvalue of $\tau_A$ is denoted by $\lambda$.
Applying Lemma~\ref{lem:convex-testing-lower} yields
\begin{equation}
\label{eq:convex-testing-profile}
\Delta_n
\geq\frac12
\left(
\frac{[A_n(r)-e^{-nr}]_+}
{K_p(|B|/\lambda^{3/2})^{n/p}
\max\{M_n^{1/p},\sqrt{M_ne^{-nr}}\}}
\right)^{p/(p-1)}.
\end{equation}
Monotonicity of the sandwiched R{\'e}nyi divergence in its order gives
\begin{equation}\label{eq:c-s-1}
\mathsf H_r^*(\rho_{AB}\|\tau_A)
\leq[r-D_1^*(\rho_{AB}\|\tau_A)]_+<r.
\end{equation}
By Theorem~\ref{thm:best-testing-exponent}, we have
\begin{equation}\label{eq:c-s-2}
A_n(r)=\exp[-n(\mathsf H_r^*(\rho_{AB}\|\tau_A)+o(1))].
\end{equation}
Combining \eqref{eq:c-s-1} and \eqref{eq:c-s-2}, we get
\begin{equation}
\lim_{n\to\infty}
-\frac1n\log[A_n(r)-e^{-nr}]_+
=\mathsf H_r^*(\rho_{AB}\|\tau_A).
\end{equation}
For fixed $p$, we also have
\begin{equation}
\lim_{n\to\infty}\frac1n
\log\max\{M_n^{1/p},\sqrt{M_ne^{-nr}}\}
=\max\left\{\frac Rp,\frac{R-r}{2}\right\}.
\end{equation}
Taking negative logarithms in~\eqref{eq:convex-testing-profile},
dividing by $n$, and then letting $n\to\infty$ therefore yields
\begin{align}\label{eq:c-s-5}
&\limsup_{n\to\infty}
-\frac1n\log\Delta_n \leq\frac{p}{p-1}\left[
\mathsf H_r^*(\rho_{AB}\|\tau_A)
+\frac{\log|B|+\frac32\log(1/\lambda)}{p}
+\max\left\{\frac Rp,\frac{R-r}{2}\right\}
\right].
\end{align}
Letting $p\to\infty$ in \eqref{eq:c-s-5} gives
\begin{equation}
\label{eq:convex-testing-exponent-bound}
\limsup_{n\to\infty}
-\frac1n\log\Delta_n
\leq \mathsf H_r^*(\rho_{AB}\|\tau_A)+\frac12[R-r]_+, \quad r>0.
\end{equation}
Take the infimum of \eqref{eq:convex-testing-exponent-bound}
over $r>0$. The scalar identity
\eqref{eq:scalar-testing-identity}, with effective rate $\ell=R$,
gives
\begin{align}
\limsup_{n\to\infty}
-\frac1n\log\Delta_n
&\leq \inf_{r>0}\left\{\mathsf H_r^*(\rho_{AB}\|\tau_A)
+\frac12[R-r]_+\right\}\notag\\
&=\max_{1\leq\alpha\leq2}
\frac{\alpha-1}{\alpha}
\left(R-D_\alpha^*(\rho_{AB}\|\tau_A)\right).
\label{eq:convex-reliability-reverse}
\end{align}

For the matching bound, Cheng and
Gao~\cite[Theorem~6, Eq.~(106), arXiv version]{ChengGao2023Convex}
give, for $1<\alpha\leq2$,
\[
\Delta_M(\rho_{AB}\|\tau_A)
\leq2^{2/\alpha-2}
\exp\left[-\frac{\alpha-1}{\alpha}
\left(\log M-D_\alpha^*(\rho_{AB}\|\tau_A)\right)\right].
\]
Apply this bound to $\rho_{AB}^{\otimes n}$ and
$\tau_A^{\otimes n}$. Tensor additivity of $D_\alpha^*(\rho_{AB}\|\tau_A)$ yields
\[
\liminf_{n\to\infty}
-\frac1n\log\Delta_n
\geq\frac{\alpha-1}{\alpha}
\left(R-D_\alpha^*(\rho_{AB}\|\tau_A)\right).
\]
Optimizing over $1<\alpha\leq2$ and including $\alpha=1$ by
continuity proves the reverse inequality to
\eqref{eq:convex-reliability-reverse}.
Thus the limit exists and~\eqref{eq:convex-reliability} follows.
If $\rho_{AB}=\tau_A\otimes\rho_B$, the splitting error is
identically zero, as stated in Theorem~\ref{thm:convex-exponents}.
\end{proof}

\section{Privacy amplification}
\label{sec:privacy-amplification}
\subsection{Problem statement and main result}
\begin{definition}[Privacy amplification]\label{def:privacy-amplification}
Let $
\rho_{XE}=\sum_{x\in\mathcal X}P_X(x)|x\rangle\langle x|_X
\otimes\rho_E^x$ be a c-q state. Alice applies a hash function
$f:\mathcal X\to\mathcal Z=\{1,\ldots,M\}$ to $X$, while the
eavesdropper retains $E$. The resulting state is
\begin{equation*}
\begin{aligned}
\mathcal R_f(\rho_{XE})
:=\sum_{x\in\mathcal X}P_X(x)|f(x)\rangle\langle f(x)|_Z
\otimes\rho_E^x
=\sum_{z\in\mathcal Z}|z\rangle\langle z|_Z
\otimes\sum_{x:\,f(x)=z}P_X(x)\rho_E^x.
\end{aligned}
\end{equation*}
Alice seeks a key that is uniform and independent of $E$ while keeping
$|\mathcal Z|$ as large as possible. The hash is drawn independently
of the source and may be announced publicly. The expected trace distance error is,
\begin{equation*}
\frac12\mathbb E_f\left\|
\mathcal R_f(\rho_{XE})
-\pi_Z\otimes\rho_E
\right\|_1.
\end{equation*}
\end{definition}

We draw $f:\mathcal X\to\mathcal Z$ from a universal$_2$ hash
family, meaning that, for all distinct $x,x'\in\mathcal X$,
\begin{equation}\label{eq:pa-universal}
\Pr_f\{f(x)=f(x')\}\leq\frac1{|\mathcal Z|}.
\end{equation}
A family is \emph{strongly universal$_2$}~\cite{shen2022strong} if every pair of
distinct inputs has independent uniform outputs, so for $x\ne x'$ and
$z,z'\in\mathcal Z$ we have
\begin{equation*}
\Pr_f\{f(x)=z,\ f(x')=z'\}=|\mathcal Z|^{-2}.
\end{equation*}
In particular, such a family has equality in
\eqref{eq:pa-universal}.
Independent random binning assigns an independent, uniformly
distributed output to each input string; it is strongly universal$_2$.

The quantum leftover hash lemma of Tomamichel, Schaffner, Smith,
and Renner~\cite{TomamichelSchaffnerSmithRenner2011} provides
one-shot security bounds for two-universal and almost
two-universal hash families against quantum side information.
Exponential security estimates were developed by
Hayashi~\cite{Hayashi2015Privacy}; Li, Yao, and
Hayashi~\cite{LiYaoHayashi2023} determined exact high-rate exponents
for purified-distance and relative-entropy security criteria.
For trace distance, Dupuis~\cite{Dupuis2023} obtained an achievable
exponent in terms of the sandwiched R\'enyi conditional entropy,
and Shen, Gao, and Cheng~\cite{shen2022strong} established a
one-shot strong converse bound for strongly universal$_2$ hashing.
Optimal second-order rates and moderate-deviation results were
obtained in~\cite{ShenGaoCheng2024}.
Regular random binning was studied
in~\cite{ShenGaoCheng2023RegularBinning}, which relates privacy
amplification to quantum soft covering and gives achievability
and strong converse bounds for constant-type sources.
That ensemble has balanced bins; the exact reliability result
below concerns the independent assignments defined above.
For related classical exponent and second-order analyses,
see~\cite{HayashiTan2017}.

In the i.i.d.\ setting, Alice and the
eavesdropper initially share $\rho_{XE}^{\otimes n}$.
For each $n$, draw $f_n:\mathcal X^n\to\mathcal Z_n=\{1,\ldots,M_n\}$
from a universal$_2$ family, with
$\lim_{n\to\infty}n^{-1}\log M_n=R\geq0$.
Write the expected trace-distance error as
\begin{equation}\label{eq:pa-block-error}
\Delta_n:=\frac12\mathbb E_{f_n}\left\|
\mathcal R_{f_n}(\rho_{XE}^{\otimes n})
-\pi_{Z_n}\otimes\rho_E^{\otimes n}
\right\|_1.
\end{equation}
For a sequence of universal\(_2\) hash ensembles, define the
reliability function and strong converse exponent by
\begin{align*}
E_{\mathtt{pa}}(\rho_{XE},R)
:=&\liminf_{n\to\infty}-\frac1n\log\Delta_n,\\
\varGamma_{\mathtt{pa}}(\rho_{XE},R)
:=&\liminf_{n\to\infty}-\frac1n\log(1-\Delta_n).
\end{align*}
The dependence on the chosen hash ensembles is suppressed in this notation.

\begin{theorem}[Universal hashing exponents]\label{thm:pa-exponents}
Let $\rho_{XE}=\sum_{x\in\mathcal X}P_X(x)|x\rangle\langle x|\otimes\rho_E^x$
be a c-q state. For each $n$, draw
$f_n:\mathcal X^n\to\mathcal Z_n=\{1,\ldots,M_n\}$ from a universal$_2$ hash
ensemble with $n^{-1}\log M_n\to R\geq0$.
Then the limit $\varGamma_{\mathtt{pa}}$ exists, and
\begin{equation}\label{eq:pa-exact-sc}
\varGamma_{\mathtt{pa}}(\rho_{XE},R)
=\max_{0\le s\le1}(1-s)\bigl(R-H_s(X|E)_\rho\bigr),
\end{equation}
and
\begin{equation}\label{eq:pa-reliability-bound}
E_{\mathtt{pa}}(\rho_{XE},R)
\ge\max_{1\le\alpha\le2}\frac{\alpha-1}{\alpha}
\bigl(H_\alpha^*(X|E)_\rho-R\bigr).
\end{equation}
For independent random binning for all sufficiently
large \(n\), the limit $E_{\mathtt{pa}}$ also exists,
and the reliability bound is attained by
Theorem~\ref{thm:pa-binning-exact}.
\end{theorem}
Thus $H(X|E)_\rho$ is the extraction threshold. For every sequence of
universal$_2$ hash ensembles, the expected error is exponentially small when
$R<H(X|E)_\rho$ and approaches one exponentially when
$R>H(X|E)_\rho$.

\subsection{The collision matrix and the direct bound}
The following matrix estimate will be used to control the
second moment under universal\(_2\) hashing.

\begin{lemma}\label{lem:pa-collision}
Let $f:\mathcal X\to\mathcal Z$ be drawn from a universal\(_2\) hash
family with $|\mathcal Z|=M$. Then the matrix
\begin{equation*}
C_{xy}=\Pr_f\{f(x)=f(y)\}-\frac1M
\end{equation*}
satisfies
\begin{equation}\label{eq:pa-collision-bound}
0\leq C\leq2(1-M^{-1})\id.
\end{equation}
\end{lemma}

\begin{proof}
For every complex vector $v=(v_x)_{x\in\mathcal X}$,
\begin{align*}
v^*Cv
=\E_f\sum_{z\in\mathcal Z}
\left|
\sum_{x\in\mathcal X}
v_x\left(1_{\{f(x)=z\}}-\frac1M\right)
\right|^2
\geq0.
\end{align*}
Thus $C$ is positive semidefinite. Moreover,
$C_{xx}=1-M^{-1}$ and
$C_{xy}\leq0$ whenever $x\ne y$.
Writing $|v|=(|v_x|)_{x\in\mathcal X}$, we obtain
\begin{align*}
v^*Cv
&=(1-M^{-1})\|v\|_2^2
+\sum_{x\ne y}C_{xy}
\operatorname{Re}(\overline{v_x}v_y)\notag\\
&\leq(1-M^{-1})\|v\|_2^2
-\sum_{x\ne y}C_{xy}|v_x||v_y|\notag\\
&=2(1-M^{-1})\|v\|_2^2
-|v|^{\mathsf T}C|v|\notag\\
&\leq2(1-M^{-1})\|v\|_2^2.
\end{align*}
The first inequality follows from $C_{xy}\leq0$ and
$\operatorname{Re}(\overline{v_x}v_y)\geq-|v_x||v_y|$;
the last follows from $C\geq0$.
Since $v$ is arbitrary, this proves
\eqref{eq:pa-collision-bound}.
\end{proof}

The collision estimate allows us to apply Dupuis's randomization bound
to universal\(_2\) hash ensembles.
\begin{lemma}[One-shot direct bound]\label{lem:pa-direct}
Let $\rho_{XE}$ be a c-q state and draw
$f:\mathcal X\to\mathcal Z$ from a universal$_2$ family with
$|\mathcal Z|=M$. Then, for every $1<\alpha\leq2$,
\begin{equation}\label{eq:pa-direct}
\frac12\E_f\left\|
\mathcal R_f(\rho_{XE})
-\frac{\id_Z}{M}\otimes\rho_E
\right\|_1
\leq\left(\frac 12\right)^{\frac{\alpha-1}{\alpha}}
\exp\left[
\frac{\alpha-1}{\alpha}\bigl(\log M-H_\alpha^*(X|E)_\rho\bigr)
\right].
\end{equation}
\end{lemma}

\begin{proof}
Define the completely randomizing map by
\begin{align*}
\mathcal U(Y_{XE})
:=\frac{\id_Z}{M}\otimes\Tr_XY_{XE}.
\end{align*}
According to \cite[Definition~4]{Dupuis2023}, the hash ensemble is
$\lambda$-randomizing if
\begin{equation}\label{eq:pa-dupuis-hypothesis}
\E_f\| (\mathcal R_f-\mathcal U)(Y_{XE})\|_2
\leq\lambda\|Y_{XE}\|_2
\end{equation}
for every finite-dimensional system $E$ and every c-q operator
\begin{align*}
Y_{XE}=\sum_{x\in\mathcal X}|x\rangle\langle x|\otimes Y_x.
\end{align*}
For a $\lambda$-randomizing hash ensemble with $\lambda>0$,
\cite[Theorem~8]{Dupuis2023} gives that for every c-q
state $\rho_{XE}$ and every $\sigma_E\in\mathcal S(E)$ satisfying
$\supp\rho_E\subseteq\supp\sigma_E$, we have
\begin{equation}\label{eq:pa-dupuis-bound}
\begin{aligned}
\E_f\| (\mathcal R_f-\mathcal U)(\rho_{XE})\|_1
\leq& 2^{\frac2\alpha-1}\lambda^{2\frac{\alpha-1}{\alpha}}
\exp\left[
\frac{\alpha-1}{\alpha}\left(\log M+
D_\alpha^*(\rho_{XE}\|\id_X\otimes\sigma_E)\right)
\right].
\end{aligned}
\end{equation}
We now verify \eqref{eq:pa-dupuis-hypothesis} with
$\lambda=\sqrt 2$. For an arbitrary family $Y=(Y_x)_{x\in\mathcal X}\subset\mathcal{B}(E)$,
define
\begin{align*}
(\mathcal T Y)_{f,z}
:=\sum_{x\in\mathcal X}
\left(1_{\{f(x)=z\}}-\frac1M\right)Y_x.
\end{align*}
Then
\begin{align*}
(\mathcal R_f-\mathcal U)(Y_{XE})
=\sum_{z\in\mathcal Z}|z\rangle\langle z|
\otimes(\mathcal T Y)_{f,z}.
\end{align*}
Since
\begin{align*}
\sum_{z\in\mathcal Z}
\left(1_{\{f(x)=z\}}-\frac1M\right)
\left(1_{\{f(y)=z\}}-\frac1M\right)
=1_{\{f(x)=f(y)\}}-\frac1M,
\end{align*}
we obtain
\begin{align*}
\E_f\| (\mathcal R_f-\mathcal U)(Y_{XE})\|_2^2
&=\E_f\sum_{z\in\mathcal Z}\|(\mathcal T Y)_{f,z}\|_2^2
\notag\\
&=\sum_{x,y\in\mathcal X}C_{xy}\Tr(Y_x^*Y_y)
\notag\\
&\leq 2\sum_{x\in\mathcal X}\|Y_x\|_2^2
=2\|Y_{XE}\|_2^2.
\end{align*}
For the inequality, apply $C\leq 2\cdot\id$ to the vector
$((Y_x)_{ab})_{x\in\mathcal X}$ for each matrix entry $(a,b)$
in an orthonormal basis of $E$, and sum over $a,b$. Then, the Cauchy--Schwarz inequality gives
\begin{align*}
\E_f\| (\mathcal R_f-\mathcal U)(Y_{XE})\|_2
\leq\left(
\E_f\| (\mathcal R_f-\mathcal U)(Y_{XE})\|_2^2
\right)^{1/2}\leq\sqrt 2\,\|Y_{XE}\|_2.
\end{align*}
Substituting $\lambda=\sqrt 2$ into
\eqref{eq:pa-dupuis-bound} gives
\begin{equation*}
\frac12\E_f\left\|
\mathcal R_f(\rho_{XE})
-\frac{\id_Z}{M}\otimes\rho_E
\right\|_1
\leq\left(\frac 12\right)^{\frac{\alpha-1}{\alpha}}
\exp\left[
\frac{\alpha-1}{\alpha}\bigl(\log M+D_\alpha^*(\rho_{XE}\|\id_X\otimes\sigma_E)\bigr)
\right].
\end{equation*}
Finally, take the infimum over the reference state $\sigma_E$ and use the definition of $H_\alpha^*(X|E)_\rho$.
This proves \eqref{eq:pa-direct}.
\end{proof}

We next deduce \eqref{eq:pa-reliability-bound} for any sequence of
universal$_2$ ensembles
$f_n:\mathcal X^n\to\mathcal Z_n$ with
$M_n=|\mathcal Z_n|$ and $n^{-1}\log M_n\to R$. Applying Lemma~\ref{lem:pa-direct}, we obtain
\begin{align*}
\frac12\E_{f_n}\left\|
\mathcal R_{f_n}(\rho_{XE}^{\otimes n})
-\pi_{Z_n}\otimes\rho_E^{\otimes n}
\right\|_1
\leq\min_{1\leq \alpha\leq 2}
\exp\left[
\frac{\alpha-1}{\alpha}\bigl(\log M_n-H_\alpha^*(X^n|E^n)_{\rho_{XE}^{\otimes n}}\bigr)
\right] \notag\\
=\min_{1\leq \alpha\leq 2}
\exp\left[
\frac{\alpha-1}{\alpha}\bigl(\log M_n-nH_\alpha^*(X|E)_{\rho_{XE}}\bigr)
\right],
\end{align*}
where the last equality follows from the additivity of $H_\alpha^*(X|E)_{\rho_{XE}}$.
Taking the negative normalized logarithm and
the lower limit gives
\begin{align*}
E_{\mathtt{pa}}(\rho_{XE},R)
\geq\max_{1\leq\alpha\leq2}
\frac{\alpha-1}{\alpha}
\left(H_\alpha^*(X|E)_\rho-R\right).
\end{align*}
This proves \eqref{eq:pa-reliability-bound}.

The collision-equality condition in
\cite[Definition~5 and Lemma~6]{Dupuis2023} is
$\Pr_f\{f(x)=f(y)\}=1/M$ for every $x\ne y$; it follows in
particular from the pairwise-uniform definition used here.
Under this condition,
$C=(1-M^{-1})\id$, so the ensemble is $1$-randomizing.
Lemma~\ref{lem:pa-collision} extends the application of Dupuis's
general randomization theorem to the usual universal$_2$
condition \eqref{eq:pa-universal}: such ensembles are
$\sqrt2$-randomizing. The resulting change in the one-shot
prefactor is independent of blocklength and therefore leaves
the achievable reliability exponent unchanged.

\subsection{The strong converse exponent}
The strong converse bound of \cite{shen2022strong} is stated
for strongly universal\(_2\) hashing. The following two-sided
testing comparison extends that bound to universal\(_2\) hashing
and gives the matching upper bound on the strong converse exponent.

\begin{lemma}[One-shot strong converse comparison]
\label{lem:pa-testing}
Let $f:\mathcal X\to\mathcal Z$ be drawn from a universal\(_2\)
hash family with $|\mathcal Z|=M\geq1$. Then
\begin{equation}\label{eq:pa-testing-universal}
\begin{aligned}
\frac14\Tr \rho_{XE}\wedge M^{-1}\id_X\otimes\rho_E\leq
1-\frac12\E_f
\norm{\mathcal R_f(\rho_{XE})
-\frac{\id_Z}{M}\otimes\rho_E}_1\leq
2\Tr \rho_{XE}\wedge 2M^{-1}\id_X\otimes\rho_E.
\end{aligned}
\end{equation}
\end{lemma}

\begin{proof}
We first prove the lower bound in \eqref{eq:pa-testing-universal}.
For each $f$ and $z$, let
\begin{align*}
\Pi_{f,z}=\left\{\sum_{x:\,f(x)=z}P_X(x)\rho_E^x-\frac{\rho_E}{M}>0\right\},
\end{align*}
and define
\begin{align*}
S_x=\E_f\left[
\Pi_{f,f(x)}-\frac1M\sum_z\Pi_{f,z}
\right],
\qquad
S=\sum_x|x\rangle\langle x|\otimes S_x.
\end{align*}
Each $S_x$ is a difference of two effects. Hence $S$ is
self-adjoint and $-\id\leq S\leq\id$. Moreover,
\begin{align}
\Tr\rho_{XE}S
&=\E_f\sum_x\Tr P_X(x)\rho_E^x\Pi_{f,f(x)}
-\frac1M\E_f\sum_z\Tr\rho_E\Pi_{f,z}\notag\\
&=\E_f\sum_z\Tr\left(\sum_{x:\,f(x)=z}P_X(x)\rho_E^x-\frac{\rho_E}{M}\right)\Pi_{f,z}\notag\\
&=\frac12\E_f
\norm{\mathcal R_f(\rho_{XE})
-\frac{\id_Z}{M}\otimes\rho_E}_1.
\label{eq:pa:testing-source}
\end{align}
Recall the map $\mathcal T$ from the proof of
Lemma~\ref{lem:pa-direct},
\begin{align*}
(\mathcal T Y)_{f,z}
=\sum_x\left(1_{\{f(x)=z\}}-\frac1M\right)Y_x.
\end{align*}
Here the input is a family $Y=(Y_x)_x$ of operators on $E$,
and the output is a family indexed by $(f,z)$.
We equip these spaces with the Hilbert--Schmidt inner products
\begin{align*}
\langle Y,Y'\rangle
=\sum_x\Tr Y_x^*Y'_x,
\qquad
\langle Z,Z'\rangle
=\E_f\sum_z\Tr Z_{f,z}^*Z'_{f,z},
\end{align*}
respectively. Since
\begin{align*}
\E_f\sum_z
\left(1_{\{f(x)=z\}}-\frac1M\right)
\left(1_{\{f(y)=z\}}-\frac1M\right)
=C_{xy},
\end{align*}
expanding the squared norm gives
\begin{align*}
\|\mathcal T Y\|_2^2
&=\E_f\sum_z
\Tr\bigl[(\mathcal T Y)_{f,z}^*
(\mathcal T Y)_{f,z}\bigr]\notag\\
&=\sum_{x,y}C_{xy}\Tr(Y_x^*Y_y)\notag\\
&\leq2\sum_x\Tr Y_x^*Y_x
=2\|Y\|_2^2.
\end{align*}
The inequality follows from $C\leq2\id$, established in
Lemma~\ref{lem:pa-collision}: apply this scalar matrix
inequality to the vector $((Y_x)_{ab})_x$ for each matrix
entry $(a,b)$ in a fixed orthonormal basis of $E$, and
then sum over $a,b$.

To identify the adjoint, let $Y=(Y_x)_x$ and
$Z=(Z_{f,z})_{f,z}$ be arbitrary operator families.
Then
\begin{align*}
\langle\mathcal T Y,Z\rangle
=\E_f\sum_z
\Tr\bigl[(\mathcal T Y)_{f,z}^*Z_{f,z}\bigr]=\sum_x\Tr Y_x^*
\left[
\E_f\sum_z
\left(1_{\{f(x)=z\}}-\frac1M\right)Z_{f,z}
\right].
\end{align*}
Consequently,
\begin{align*}
(\mathcal T^*Z)_x
=\E_f\sum_z
\left(1_{\{f(x)=z\}}-\frac1M\right)Z_{f,z},
\qquad
\|\mathcal T^*\|_{2\to2}
=\|\mathcal T\|_{2\to2}\leq\sqrt2.
\end{align*}

Applying this formula to the family
$(\Pi_{f,z}\rho_E^{1/2})_{f,z}$ yields
\begin{align*}
\left[
\mathcal T^*
\bigl((\Pi_{f,z}\rho_E^{1/2})_{f,z}\bigr)
\right]_x
&=\E_f\sum_z
\left(1_{\{f(x)=z\}}-\frac1M\right)
\Pi_{f,z}\rho_E^{1/2}\notag\\
&=\left(
\E_f\Pi_{f,f(x)}
-\frac1M\E_f\sum_z\Pi_{f,z}
\right)\rho_E^{1/2}\notag\\
&=S_x\rho_E^{1/2}.
\end{align*}
The second equality uses linearity and the fact that
$\rho_E^{1/2}$ is independent of $f$ and $z$; the last
equality is the definition of $S_x$.
The adjoint norm bound therefore gives
\begin{align}
\sum_x\Tr\rho_ES_x^2
&=\sum_x\|S_x\rho_E^{1/2}\|_2^2\notag\\
&=\left\|
\mathcal T^*\bigl((\Pi_{f,z}\rho_E^{1/2})_{f,z}\bigr)
\right\|_2^2\notag\\
&\leq2\E_f\sum_z\|\Pi_{f,z}\rho_E^{1/2}\|_2^2\notag\\
&=2\E_f\sum_z\Tr\rho_E\Pi_{f,z}.
\label{eq:pa:second-moment}
\end{align}
Here the final equality follows from $\Pi_{f,z}^2=\Pi_{f,z}$.
The two error costs of the optimal output test satisfy
\begin{align*}
1-\frac12\E_f
\norm{\mathcal R_f(\rho_{XE})
-\frac{\id_Z}{M}\otimes\rho_E}_1
=&
\E_f\sum_z\Tr \sum_{x:f(x)=z}P_X(x)\rho_E^x(\id-\Pi_{f,z})
+\frac1M\E_f\sum_z\Tr\rho_E\Pi_{f,z}\notag\\
\geq&\frac1M\E_f\sum_z\Tr\rho_E\Pi_{f,z}.
\end{align*}
Consequently, \eqref{eq:pa:second-moment} gives
\begin{align*}
M^{-1}\Tr\bigl[(\id_X\otimes\rho_E)S^2\bigr]
=\frac1M\sum_x\Tr\rho_ES_x^2\leq
2\left(
1-\frac12\E_f
\norm{\mathcal R_f(\rho_{XE})
-\frac{\id_Z}{M}\otimes\rho_E}_1
\right).
\end{align*}
Together with \eqref{eq:pa:testing-source} and
$-\id\leq S\leq\id$, this verifies the hypotheses of
Lemma~\ref{lem:test-transfer} with
$\rho=\rho_{XE}$, $\sigma=\id_X\otimes\rho_E$,
$m=M^{-1}$, $K=2$, and
\begin{align*}
\Delta=\frac12\E_f
\norm{\mathcal R_f(\rho_{XE})
-\frac{\id_Z}{M}\otimes\rho_E}_1\in[0,1].
\end{align*}
Therefore, Lemma~\ref{lem:test-transfer}
yields the lower bound in \eqref{eq:pa-testing-universal}.

For the upper bound in \eqref{eq:pa-testing-universal},
we evaluate the binary testing cost using the effects
\begin{align*}
\begin{aligned}
T_z(f)=
\left(
\sum_{x:\,f(x)=z}P_X(x)\rho_E^x
+\frac{\rho_E}{M}
\right)^{-1/2}
\sum_{x:\,f(x)=z}P_X(x)\rho_E^x
\left(
\sum_{x:\,f(x)=z}P_X(x)\rho_E^x
+\frac{\rho_E}{M}
\right)^{-1/2}.
\end{aligned}
\end{align*}
Since $0\leq T_z(f)\leq\id$ for every $z,f$, the pair
$(T_z(f),\id-T_z(f))$ is a binary measurement. Consequently,
\begin{align*}
&1-\frac12\E_f
\norm{\mathcal R_f(\rho_{XE})
-\frac{\id_Z}{M}\otimes\rho_E}_1 \notag\\
=&\E_f\sum_z
\Tr
\left(\sum_{x:f(x)=z}P_X(x)\rho_E^x\right)\wedge
\frac{\rho_E}{M} \notag\\
\leq&\E_f\sum_z\left(\Tr\sum_{x:f(x)=z}P_X(x)\rho_E^x(\id-T_z(f))+ \Tr \frac{\rho_E}{M}T_z(f) \right)   \notag\\
=&
2\E_f\sum_x
\Tr P_X(x)\rho_E^x
\left(
\sum_{x':\,f(x')=f(x)}P_X(x')\rho_E^{x'}
+\frac{\rho_E}{M}
\right)^{-1/2}
\frac{\rho_E}{M}
\left(
\sum_{x':\,f(x')=f(x)}P_X(x')\rho_E^{x'}
+\frac{\rho_E}{M}
\right)^{-1/2}
\notag\\
\leq&
2\E_f\sum_x
\Tr
P_X(x)\rho_E^x \wedge
\left(\sum_{x'\ne x}
1_{\{f(x')=f(x)\}}P_X(x')\rho_E^{x'}
+\frac{\rho_E}{M}
\right).
\end{align*}
The first inequality follows from \eqref{eq:testing-cost}. The second
equality uses the identity
\begin{align*}
\Tr\left[
\left(
\sum_{x:\,f(x)=z}P_X(x)\rho_E^x
\right)
(\id-T_z(f))
\right]
=
\Tr\left[\frac{\rho_E}{M}T_z(f)\right],
\end{align*}
which holds for every \(f\) and \(z\). For the last inequality, first replace the middle factor
$\rho_E/M$ by
\begin{align*}
\sum_{x'\ne x}
1_{\{f(x')=f(x)\}}P_X(x')\rho_E^{x'}
+\frac{\rho_E}{M},
\end{align*}
keeping both inverse factors unchanged, and then apply
Lemma~\ref{lem:pgm} to $
P_X(x)\rho_E^x$ and $
\sum_{x'\ne x}
1_{\{f(x')=f(x)\}}P_X(x')\rho_E^{x'}
+\frac{\rho_E}{M}.$
By the universal\(_2\) collision bound, we get
\begin{align*}
\E_f\left[
\sum_{x'\ne x}
1_{\{f(x')=f(x)\}}P_X(x')\rho_E^{x'}
\right]
&=
\sum_{x'\ne x}
\Pr_f\{f(x')=f(x)\}P_X(x')\rho_E^{x'}\notag\\
&\leq
\frac1M\left(\rho_E-P_X(x)\rho_E^x\right)
\leq\frac{\rho_E}{M}.
\end{align*}
By the joint concavity and monotonicity of
$(A,B)\mapsto\Tr A\wedge B$, we have
\begin{align*}
&\E_f\sum_x
\Tr
P_X(x)\rho_E^x\wedge
\left(\sum_{x'\ne x}
1_{\{f(x')=f(x)\}}P_X(x')\rho_E^{x'}
+\frac{\rho_E}{M}
\right)
\notag\\
\leq&
\sum_x
\Tr
P_X(x)\rho_E^x\wedge
\left(\E_f\left[
\sum_{x'\ne x}
1_{\{f(x')=f(x)\}}P_X(x')\rho_E^{x'}
\right]
+\frac{\rho_E}{M}
\right)
\notag\\
\leq&
\sum_x
\Tr
P_X(x)\rho_E^x\wedge
\frac{2\rho_E}{M}
\notag\\
=&
\Tr
\rho_{XE}\wedge
\frac2M\id_X\otimes\rho_E.
\end{align*}
The last equality is the direct-sum rule for binary testing.
This proves the upper bound in \eqref{eq:pa-testing-universal}.
\end{proof}
To apply Lemma~\ref{lem:weighted-testing}, take
$\rho=\rho_{XE}$, $\sigma=\pi_X\otimes\rho_E$, and
$m_n=|\mathcal X|^n/|\mathcal Z_n|$. The weight satisfies
$n^{-1}\log m_n\to\log|\mathcal X|-R$.
The operators $\rho$ and $\sigma$ are density operators, and
\begin{align*}
\supp\rho_{XE}
\subseteq\supp(\id_X\otimes\rho_E)
=\supp(\pi_X\otimes\rho_E).
\end{align*}
Hence Lemma~\ref{lem:weighted-testing} gives
\begin{align*}
\lim_{n\to\infty}-\frac1n\log
\Tr \rho_{XE}^{\otimes n}\wedge
|\mathcal Z_n|^{-1}\id_{X^n}\otimes\rho_E^{\otimes n}=
\max_{0\leq s\leq1}(1-s)
\bigl(R-H_s(X|E)_\rho\bigr).
\end{align*}
The fixed prefactors and the factor of two between the testing weights
in \eqref{eq:pa-testing-universal} do not affect the exponent.
Applying that comparison to
$\rho_{XE}^{\otimes n}$ with
$n^{-1}\log|\mathcal Z_n|\to R$, we obtain
\begin{align*}
\varGamma_{\mathtt{pa}}(\rho_{XE},R)
&=\lim_{n\to\infty}-\frac1n\log(1-\Delta_n)\notag\\
&=\lim_{n\to\infty}-\frac1n\log
\Tr \rho_{XE}^{\otimes n}\wedge
|\mathcal Z_n|^{-1}\id_{X^n}\otimes\rho_E^{\otimes n} \notag\\
&=\max_{0\leq s\leq1}(1-s)
\bigl(R-H_s(X|E)_\rho\bigr).
\end{align*}
This proves \eqref{eq:pa-exact-sc} for every sequence of
universal\(_2\) hash ensembles.

\subsection{The reliability function for random binning}
Here $f$ is drawn from the independent random-binning ensemble with
$M\geq2$.
\begin{lemma}
\label{lem:pa-testing-lower}
Let $M\geq2$ and draw $f:\mathcal X\to\mathcal{Z}=\{1,\ldots,M\}$
by independent random binning. Suppose that $0\leq T_x\leq\id_E$
satisfy
\begin{equation*}
\frac{1}{|\mathcal{X}|}\sum_xT_x\leq b\id_E.
\end{equation*}
Then, for every $p\geq2$,
\begin{equation*}
\frac{1}{2} \mathbb{E}_f \norm{\mathcal{R}_{f}(\rho_{XE})-\frac{\id_Z}{|\mathcal{Z}|}\otimes \rho_E}_1
\geq
\frac12
\left(
\frac{\E_X \Tr(\rho_E^XT_X)}
{2K_p(M|E|)^{1/p}
 \max\{1,\sqrt{|\mathcal X|b/M}\}}
\right)^{p/(p-1)},
\end{equation*}
where $K_p$ is the constant in
Lemma~\ref{lem:rosenthal-moments}.
\end{lemma}
\begin{proof}
We apply Lemma~\ref{general-one-shot} with
\[
\mathcal M
=L_\infty(\mathcal Z^{\mathcal X})
\overline\otimes\mathcal B(ZE),
\qquad
\mathcal N=\mathcal B(ZE),
\qquad
\mathcal E=\E_f,
\]
and
\[
S=\sum_{z=1}^M|z\rangle\langle z|_Z
\otimes\sum_x1_{\{f(x)=z\}}T_x,
\qquad
\rho=\mathcal R_f(\rho_{XE}),
\qquad
\varphi=\frac{\id_{ZE}}{M|E|}.
\]
Here $\mathcal Z^{\mathcal X}$ carries the uniform measure.
The trace of $\mathcal M$ is $\E_f\Tr$ and $\E_f$ preserves $\varphi$,
with
\[
\E_f\mathcal R_f(\rho_{XE})
=\frac{\id_Z}{M}\otimes\rho_E.
\]
Since
\[
-\id_{ZE}
\leq
\mathcal R_f(\rho_{XE})-\frac{\id_Z}{M}\otimes\rho_E
\leq
\id_{ZE},
\]
the boundedness assumption in Lemma~\ref{general-one-shot}
holds with $L=M|E|$. Hence we get
\begin{equation}
\label{pa-one}
\left|\E_f\Tr\bigl[
\mathcal R_f(\rho_{XE})(S-\E_fS)
\bigr]\right|
\leq
(M|E|)^{1/p}
\left(\E_f\left\|
\mathcal R_f(\rho_{XE})-\frac{\id_Z}{M}\otimes\rho_E
\right\|_1\right)^{1-1/p}
\|S-\E_fS\|_{p,\varphi}.
\end{equation}

For the numerator, independence of the values $f(x)$ gives
\[
\E_f\Tr\bigl[
\mathcal R_f(\rho_{XE})(S-\E_fS)
\bigr]
=
\left(1-\frac1M\right)\E_X\Tr(\rho_E^XT_X)
\geq
\frac12\E_X\Tr(\rho_E^XT_X).
\]
For the centered norm, for each fixed $z$, the matrices
$(1_{\{f(x)=z\}}-1/M)T_x$ are independent, have expectation
zero, and have operator norm at most one. Thus, for $p\geq2$,
we have
\begin{align*}
\left(
\sum_x\E_f
\Tr\left|
\left(1_{\{f(x)=z\}}-\frac1M\right)T_x
\right|^p
\right)^{1/p}\leq
\left(
\frac1M\left(1-\frac1M\right)
\Tr\sum_xT_x^2
\right)^{1/p}\leq
\left(\frac{|\mathcal X||E|b}{M}\right)^{1/p},
\end{align*}
and
\begin{align*}
\left[
\Tr\left(
\sum_x\E_f
\left(
\left(1_{\{f(x)=z\}}-\frac1M\right)T_x
\right)^2
\right)^{p/2}
\right]^{1/p}
&=
\sqrt{\frac1M\left(1-\frac1M\right)}
\left[\Tr\left(\sum_xT_x^2\right)^{p/2}\right]^{1/p}\\
&\leq
|E|^{1/p}\sqrt{\frac{|\mathcal X|b}{M}}.
\end{align*}
Here we used $0\leq T_x\leq\id_E$ and
$0\leq\sum_xT_x^2\leq\sum_xT_x\leq|\mathcal X|b\id_E$.
Applying Lemma~\ref{lem:rosenthal-moments} to
$(1_{\{f(x)=z\}}-1/M)T_x$, $x\in\mathcal X$ yields
\begin{align*}
\|S-\E_fS\|_{p,\varphi}
&=
\left(
\frac{1}{M|E|}
\sum_{z=1}^M\E_f\Tr
\left|
\sum_x\left(1_{\{f(x)=z\}}-\frac1M\right)T_x
\right|^p
\right)^{1/p}\\
&\leq
K_p\max\left\{
\left(\frac{|\mathcal X|b}{M}\right)^{1/p},
\sqrt{\frac{|\mathcal X|b}{M}}
\right\}\\
&\leq
K_p\max\left\{1,\sqrt{\frac{|\mathcal X|b}{M}}\right\}.
\end{align*}
If $\E_X\Tr(\rho_E^XT_X)=0$, the asserted lower bound is
immediate. Otherwise, substituting the numerator and norm bounds
into \eqref{pa-one} and rearranging gives
\begin{equation*}
\frac12\E_f\left\|
\mathcal R_f(\rho_{XE})-\frac{\id_Z}{M}\otimes\rho_E
\right\|_1
\geq\frac12
\left(
\frac{\E_X\Tr(\rho_E^XT_X)}
{2K_p(M|E|)^{1/p}
\max\{1,\sqrt{|\mathcal X|b/M}\}}
\right)^{p/(p-1)},
\end{equation*}
as claimed.
\end{proof}

\begin{theorem}[Random-binning reliability]\label{thm:pa-binning-exact}
Let $\rho_{XE}$ be a fixed c-q state. For each $n$, draw
$f_n:\mathcal X^n\to\mathcal Z_n$ by independent random binning. Assume
$M_n=|\mathcal{Z}_n|\ge2$ for all sufficiently large \(n\) and
$n^{-1}\log M_n\to R\ge0$. Then
\begin{equation}\label{eq:pa-binning-exact}
\lim_{n\to\infty}-\frac1n\log\Delta_n
 =\max_{1\le\alpha\le2}\frac{\alpha-1}{\alpha}
  \bigl(H_\alpha^*(X|E)_\rho-R\bigr).
\end{equation}
At any blocklength with $M_n=1$, the error is zero.
\end{theorem}
\begin{proof}
We first establish the lower bound on the reliability exponent.
Since the random-binning ensemble is universal$_2$,
Lemma~\ref{lem:pa-direct} implies that for every fixed $1<\alpha\leq2$ we have
\begin{equation*}
\Delta_n
\leq
2^{-(\alpha-1)/\alpha}
\exp\left[
-\frac{\alpha-1}{\alpha}
\left(
nH_\alpha^*(X|E)_\rho-\log M_n
\right)
\right],
\end{equation*}
where we have used the additivity of
$H_\alpha^*(X|E)_\rho$.
Taking negative logarithms, dividing by $n$ and using
$n^{-1}\log M_n\to R$, we obtain
\begin{equation*}
\liminf_{n\to\infty}-\frac1n\log\Delta_n
\geq
\max_{1\leq\alpha\leq2}
\frac{\alpha-1}{\alpha}
\left(H_\alpha^*(X|E)_\rho-R\right).
\end{equation*}

For the matching upper bound on the reliability exponent, we apply
Lemma~\ref{lem:pa-testing-lower} to an optimal conditional test.
The testing pair in Theorem~\ref{thm:best-testing-exponent} is
$(\rho_{XE},\pi_X)$; write $A_n(r)$ for its optimal success
probability.
The support condition in that theorem is automatic because
$\pi_X$ is faithful.
Since $\pi_X=\id_X/|\mathcal X|$, scaling the second argument
of the sandwiched R\'enyi divergence gives
\begin{equation*}
D_\alpha^*(\rho_{XE}\|\pi_X)
=
\log|\mathcal X|-H_\alpha^*(X|E)_\rho,
\qquad \alpha>1.
\end{equation*}

Fix $r\geq0$.
The feasible set in the definition of $A_n(r)$ is compact and
contains $e^{-nr}\id_{X^nE^n}$.
Consequently, the maximum is attained and
\begin{equation*}
A_n(r)\geq e^{-nr}>0.
\end{equation*}
Dephasing a maximizing test in the classical basis of $X^n$
preserves the inequalities $0\leq T_n\leq\id_{X^nE^n}$.
It also leaves its partial trace over $X^n$ unchanged.
Since $\pi_X^{\otimes n}=|\mathcal X|^{-n}\id_{X^n}$, the
conditional constraint is therefore preserved.
Finally, $\rho_{XE}^{\otimes n}$ is diagonal in the same
classical basis, so dephasing preserves the objective value.
We may thus choose a maximizing test of the form
\begin{equation*}
T_n
=
\sum_{x^n\in\mathcal X^n}
|x^n\rangle\langle x^n|\otimes T_{n,x^n},
\qquad
0\leq T_{n,x^n}\leq\id_{E^n},
\end{equation*}
satisfying
\begin{align*}
\sum_{x^n\in\mathcal X^n}
P_X^{\otimes n}(x^n)
\Tr\!\left(\rho_{E^n}^{x^n}T_{n,x^n}\right)=A_n(r),\qquad
\frac1{|\mathcal X|^n}
\sum_{x^n\in\mathcal X^n}T_{n,x^n}
\leq e^{-nr}\id_{E^n}.
\end{align*}

Fix $p\geq2$. For sufficiently large $n$, we have $M_n\geq2$.
Apply Lemma~\ref{lem:pa-testing-lower} to
$\rho_{XE}^{\otimes n}$ and the family
$\{T_{n,x^n}\}_{x^n\in\mathcal X^n}$, with
$M=M_n$ and $b=e^{-nr}$. This gives
\begin{equation}
\Delta_n
\geq
\frac12
\left(
\frac{A_n(r)}
{2K_p(M_n|E|^n)^{1/p}
 \max\left\{1,\sqrt{|\mathcal X|^n e^{-nr}/M_n}\right\}}
\right)^{p/(p-1)}.
\end{equation}
Here $K_p$ depends only on $p$. Since $p$ is fixed, it holds that $\log(2K_p)/n\to0$.
By Theorem~\ref{thm:best-testing-exponent}, we have
\[
-\frac1n\log A_n(r)
\longrightarrow \mathsf H_r^*(\rho_{XE}\|\pi_X).
\]
Using $n^{-1}\log M_n\to R$ and continuity of the
positive-part function, we conclude that
\begin{align*}
\limsup_{n\to\infty}-\frac1n\log\Delta_n
\leq
\frac{p}{p-1}
\left[
\mathsf H_r^*(\rho_{XE}\|\pi_X)
+\frac{R+\log|E|}{p}
+\frac12[\log|\mathcal X|-R-r]_+
\right].
\end{align*}
This holds for every $p\geq2$. Letting $p\to\infty$ gives,
for every $r\geq0$,
\begin{equation}\label{eq:pa-testing-exponent-bound}
\limsup_{n\to\infty}-\frac1n\log\Delta_n
\leq
\mathsf H_r^*(\rho_{XE}\|\pi_X)
+\frac12[\log|\mathcal X|-R-r]_+.
\end{equation}
We optimize \eqref{eq:pa-testing-exponent-bound} over $r>0$.
The shared identity \eqref{eq:scalar-testing-identity} applies with
the testing pair $(\rho_{XE},\pi_X)$ and real effective rate
$\log|\mathcal X|-R$. Using
$D_\alpha^*(\rho_{XE}\|\pi_X)
=\log|\mathcal X|-H_\alpha^*(X|E)_\rho$, it gives
\begin{align*}
\limsup_{n\to\infty}-\frac1n\log\Delta_n
&\leq
\inf_{r>0}
\left\{
\mathsf H_r^*(\rho_{XE}\|\pi_X)
+\frac12[\log|\mathcal X|-R-r]_+
\right\}\\
&=\max_{1\leq\alpha\leq2}
\frac{\alpha-1}{\alpha}
\left(\log|\mathcal X|-R
-D_\alpha^*(\rho_{XE}\|\pi_X)\right)\\
&=
\max_{1\leq\alpha\leq2}
\frac{\alpha-1}{\alpha}
\left(H_\alpha^*(X|E)_\rho-R\right).
\end{align*}
Combining this upper bound with the lower bound proved above
shows that the limit exists and establishes
\eqref{eq:pa-binning-exact}.
\end{proof}

\subsection{Universality does not determine the reliability exponent}
The reliability exponent can depend on the hash ensemble even within
the class of strongly universal\(_2\) families. A universal$_2$ family
can also achieve zero error for the same source.

Take $E$ to be trivial and $X$ uniform on $\mathbb F_q$, for a fixed
prime power $q$. For each $n$, choose an $\mathbb F_q$-linear
identification $\mathcal X^n=\mathbb F_q^n\cong\mathbb F_{q^n}$.
Thus $X^n$ is uniform on $\mathbb F_{q^n}$ and the source is i.i.d.
Choose integers $1\leq m_n\leq n$ such that
\[
M_n=q^{m_n},
\qquad
\frac{m_n\log q}{n}\longrightarrow R\in(0,\log q).
\]
Let
\[
\pi_{m_n}:\mathbb F_{q^n}\longrightarrow\mathbb F_q^{m_n}
\]
be the projection onto the first $m_n$ coordinates under this
identification.
We identify the output alphabet with $\mathbb F_q^{m_n}$.
For each ensemble considered below, $\Delta_n$ denotes the expected
total variation distance of the output distribution from the
uniform distribution. The distance is evaluated for each fixed
hash function before averaging over the random seed.

First, choose $a,b$ independently and uniformly from
$\mathbb F_{q^n}$, and define
\[
h_{a,b}(x)=\pi_{m_n}(ax+b).
\]
For any distinct $x,y\in\mathbb F_{q^n}$, the map
\[
(a,b)\longmapsto(ax+b,ay+b)
\]
is a bijection of $\mathbb F_{q^n}^2$, since $x-y$ is invertible.
Consequently, $(ax+b,ay+b)$ is uniformly distributed on
$\mathbb F_{q^n}^2$. Applying the coordinate projection to both
components gives
\[
\Pr_{a,b}\{h_{a,b}(x)=z,\ h_{a,b}(y)=z'\}
=\frac1{M_n^2}
\]
for every $z,z'\in\mathbb F_q^{m_n}$.
Thus this ensemble is strongly universal$_2$.

If $a\ne0$, the map $x\mapsto ax+b$ is a permutation of
$\mathbb F_{q^n}$. Since every fiber of $\pi_{m_n}$ has
cardinality $q^{n-m_n}$, the output of the uniform source is
exactly uniform on $\mathbb F_q^{m_n}$.
If $a=0$, the output is the fixed value $\pi_{m_n}(b)$, whose
total variation distance from the uniform distribution is
$1-M_n^{-1}$.
Since $\Pr(a=0)=q^{-n}$, we obtain
\begin{equation}
\label{eq:pa-field-example}
\begin{aligned}
\Delta_n=q^{-n}(1-M_n^{-1}),\qquad
\lim_{n\to\infty}-\frac1n\log\Delta_n=\log q.
\end{aligned}
\end{equation}
Here we used $M_n\to\infty$, which follows from $R>0$.

We next consider independent random binning with the same input and
output alphabets. This ensemble is also strongly universal$_2$.
Write $N=q^n$. The number of inputs assigned to any fixed output
bin has distribution
\[
K\sim\operatorname{Bin}(N,M_n^{-1}).
\]
For each realization of the hash function, the probability of
that output is its bin occupancy divided by $N$.
By symmetry of the output bins,
\[
\Delta_n
=
\frac{M_n}{2N}\E\left|K-\frac{N}{M_n}\right|.
\]
Set
\[
v=\frac{N}{M_n}\left(1-\frac1{M_n}\right).
\]
The binomial central moments satisfy
\[
\E(K-\E K)^2=v
\]
and
\begin{align*}
\E(K-\E K)^4
=
3v^2+
v\left[
1-\frac6{M_n}\left(1-\frac1{M_n}\right)
\right]
\leq3v^2+v.
\end{align*}
Applying H\"older's inequality with conjugate exponents
$3/2$ and $3$ gives
\begin{align*}
v
=
\E\left[
|K-\E K|^{2/3}|K-\E K|^{4/3}
\right]
\leq
\left(\E|K-\E K|\right)^{2/3}
\left(\E|K-\E K|^4\right)^{1/3}.
\end{align*}
Combining this estimate with the fourth-moment bound and the
Cauchy--Schwarz inequality yields
\[
\frac{v^{3/2}}{\sqrt{3v^2+v}}
\leq
\E|K-\E K|
\leq
\sqrt v.
\]
Therefore,
\[
\frac{\sqrt{1-M_n^{-1}}}{2\sqrt{3+v^{-1}}}
\sqrt{\frac{M_n}{N}}
\leq
\Delta_n
\leq
\frac{\sqrt{1-M_n^{-1}}}{2}
\sqrt{\frac{M_n}{N}}.
\]
The rate assumption implies both $M_n\to\infty$ and
$N/M_n\to\infty$, and hence $v\to\infty$.
The prefactors in these bounds consequently converge to strictly
positive constants. Taking negative logarithms and dividing by
$n$, we conclude that
\[
\lim_{n\to\infty}-\frac1n\log\Delta_n
=
\frac{\log q-R}{2}.
\]
This differs strictly from the exponent in
\eqref{eq:pa-field-example}, although both ensembles are strongly
universal$_2$.

Finally, restrict the affine hash ensemble to $a\ne0$.
Every hash function in the resulting ensemble maps the uniform
source to the uniform output distribution, so
\(
\Delta_n=0
\) for every $n$.
Nevertheless, the ensemble remains universal$_2$.
Indeed, for distinct $x,y$, the random variable $a(x-y)$ is
uniformly distributed over the nonzero elements of
$\mathbb F_{q^n}$. The kernel of $\pi_{m_n}$ contains
$q^{n-m_n}-1$ nonzero elements. Thus
\begin{align*}
\Pr\{h_{a,b}(x)=h_{a,b}(y)\mid a\ne0\}
&=
\Pr\{\pi_{m_n}(a(x-y))=0\mid a\ne0\}\\
&=
\frac{q^{n-m_n}-1}{q^n-1}\\
&\leq q^{-m_n}
=\frac1{M_n}.
\end{align*}
Under the convention $-\log0=+\infty$, this ensemble therefore
has reliability exponent $+\infty$ for the specified source.

These hash ensembles show that universality alone does not
determine the exact reliability exponent. In particular, the
first two ensembles have identical pairwise output distributions,
and hence identical second moments for their bin occupancy
counts, but different exponential rates of decay for the expected
total variation distance. For the first affine ensemble, the average
error is entirely due to the rare event $a=0$, whereas independent
random binning produces fluctuations whose expected absolute
size is comparable to their standard deviation.

Consequently, the independence of the hash outputs in
Theorem~\ref{thm:pa-binning-exact} cannot be replaced by strong
universality without additional assumptions controlling
fluctuations beyond their second moments.
These examples are consistent with
\eqref{eq:pa-reliability-bound}: the general direct bound for
universal$_2$ hashing provides a lower bound on the reliability
exponent, but need not be tight for a particular hash ensemble.

\section{Partial-trace decoupling}
\label{sec:partial-trace}
\subsection{Problem statement and main result}
\begin{definition}[Standard partial-trace decoupling]
\label{def:partial-trace-decoupling}
Let $\rho_{AE}$ be a bipartite quantum state, and let $A=C\otimes D$ with
$c:=|C|\geq2$, $k:=|D|\geq1$, and $d:=|A|=ck$.
Alice applies a unitary $U$ drawn from normalized Haar measure on the
unitary group of $A$ and traces out $D$, aiming to leave $C$ maximally
mixed and independent of $E$. Write
\[
\rho_{CE}^{U}:=\operatorname{Tr}_D\left[
(U\otimes\id_E)\rho_{AE}(U^*\otimes\id_E)
\right]
\]
for the resulting state on $CE$.
The expected trace-distance error is
\begin{equation}\label{eq:dec:error}
\frac12\mathbb E_U\left\|
\rho_{CE}^{U}
-\pi_C\otimes\rho_E
\right\|_1.
\end{equation}
\end{definition}

One-shot decoupling for Haar-random unitaries followed by a general
quantum channel was established in terms of smooth conditional
entropies by Dupuis, Berta, Wullschleger, and
Renner~\cite{DupuisBertaWullschlegerRenner2014}.
Dupuis~\cite{Dupuis2023} derived R\'enyi entropy bounds for
decoupling without smoothing.
Cheng, Dupuis, and Gao~\cite{ChengDupuisGao2024} obtained sandwiched
R\'enyi achievability and Petz R\'enyi strong converse bounds
directly in trace distance.
More recently, Berta, Cheng, and Yao~\cite{BertaChengYao2026}
characterized the Haar partial-trace error exponent under quantum
relative entropy in the low-cost-rate regime.
We determine the exact trace-distance exponents for the i.i.d.\
partial-trace model below.

In the i.i.d.\ setting, the single-copy state $\rho_{AE}$ and
factorization $A=C\otimes D$ are fixed.
At blocklength $n$, Alice applies a Haar-distributed unitary $U_n$
on the full unitary group of $A^n\cong C^nD^n$ and traces out $D^n$.
The expected decoupling error is
\begin{equation}
\label{eq:dec:block-error}
\Delta_n:=\frac12\E_{U_n}
\left\|\rho_{C^nE^n}^{U_n}
-\pi_C^{\otimes n}\otimes\rho_E^{\otimes n}\right\|_1,
\end{equation}
where $\rho_{C^nE^n}^{U_n}
:=\Tr_{D^n}\left[
(U_n\otimes \id_{E^n})\rho_{AE}^{\otimes n}
(U_n^*\otimes \id_{E^n})\right]$.
Define the reliability function and strong converse exponent by
\begin{align*}
E_{\mathtt{dec}}(\rho_{AE},c,k)
&:=\liminf_{n\to\infty}-\frac1n\log\Delta_n,\\
\varGamma_{\mathtt{dec}}(\rho_{AE},c,k)
&:=\liminf_{n\to\infty}-\frac1n\log(1-\Delta_n).
\end{align*}

\begin{theorem}[Exact exponents for partial trace]
\label{thm:dec:exact}
Let $\rho_{AE}$ be a  bipartite state, and let $A=C\otimes D$
with $c=|C|\geq2$ and $k=|D|\geq1$. 
The limit $E_{\mathtt{dec}}$ and $\varGamma_{\mathtt{dec}}$
exist, and
\begin{equation}
\varGamma_{\mathtt{dec}}(\rho_{AE},c,k)
=\max_{0\leq s\leq1}(1-s)\left\{\log\frac{c}{k}-H_s(A|E)_\rho\right\},
\label{eq:dec:strong}
\end{equation}
\begin{equation}
E_{\mathtt{dec}}(\rho_{AE},c,k)
  =\max_{1\leq\alpha\leq2}\frac{\alpha-1}{\alpha}
\left\{H_\alpha^*(A|E)_\rho-\log\frac{c}{k}\right\}.
 \label{eq:dec:reliability}
\end{equation}
\end{theorem}
For $\rho_{AE}\ne\pi_A\otimes\rho_E$, the expected error decays
exponentially when $\log(c/k)<H(A|E)_\rho$ and approaches one
exponentially when $\log(c/k)>H(A|E)_\rho$. Thus the ratio of retained
to discarded dimensions determines which regime applies.

The conditional entropies are those in \eqref{eq:entropies}.
We work on $\supp\rho_E$; the joint state $\rho_{AE}$ may be singular.
The one-shot estimates of Cheng, Dupuis, and Gao
\cite[Theorem~2, Corollary~12, and Example~13 of the arXiv version]{ChengDupuisGao2024},
specialized to partial trace, give
\begin{align}
\frac12\mathbb E_U\left\|
\rho_{CE}^{U}
-\pi_C\otimes\rho_E
\right\|_1&\leq C_0
\exp\left\{-\frac{\alpha-1}{\alpha}
\left[H_\alpha^*(A|E)_\rho-\log\frac{c}{k}\right]\right\},
&&1<\alpha\leq2,\label{eq:dec:known-error}\\
 1-\frac12\mathbb E_U\left\|
 \rho_{CE}^{U}
 -\pi_C\otimes\rho_E
 \right\|_1&\leq C_0
\exp\left\{-(1-s)
\left[\log\frac{c}{k}-H_s(A|E)_\rho\right]\right\},
&&0<s<1,\label{eq:dec:known-strong}
\end{align}
where $C_0$ is an absolute constant. The normalized Choi state of the
partial trace is $\Phi_{C'C}\otimes\pi_{D'}$, where $C'D'$ is a copy
of the input $CD$ and $\Phi_{C'C}$ is maximally entangled. The Choi state
is invariant under $\overline V_{C'}\otimes V_C$ for every unitary $V_C$.
Averaging over these unitaries sends any conditioning state on $C$ to
$\pi_C$, so data processing reduces the sandwiched optimization to that
reference. Both conditional R\'enyi entropies of the Choi state equal
$\log k-\log c$.

\subsection{The strong converse exponent}
\begin{proof}[Proof of \eqref{eq:dec:strong}]
We first prove the one-shot comparison
\begin{equation}\label{eq:dec:strong-reverse}
\frac13\Tr[\rho_{AE}\wedge k^2(\pi_A\otimes\rho_E)]
\leq 1-\Delta,
\qquad
\Delta:=\frac12\E_U\norm{\rho_{CE}^{U}-\pi_C\otimes\rho_E}_1.
\end{equation}
We use Lemma~\ref{lem:test-transfer} with $\rho=\rho_{AE}$,
$\sigma=\pi_A\otimes\rho_E$, $m=k^2$, and $K=1$.
Define the trace-preserving conditional expectations
$\mathcal N,\mathcal P_U:\mathcal B(AE)\to\mathcal B(AE)$ by
\begin{equation}
\label{eq:dec:conditional-expectation}
\mathcal N(X)=\pi_A\otimes\Tr_A X,
\qquad
\mathcal P_U(X)
=U^*\left[\frac{1}{k}\Tr_D(UXU^*)\otimes\id_D\right]U.
\end{equation}
Here and below, unitary conjugations act trivially on $E$,
and the tensor factors are placed according to their register
labels. We use these maps throughout this section. They satisfy
\begin{align*}
\mathcal N^2=\mathcal N,\qquad
\mathcal P_U^2=\mathcal P_U,\qquad
\mathcal P_U\mathcal N=\mathcal N\mathcal P_U=\mathcal N.
\end{align*}
Writing $d=|A|=ck$, the Haar-twirling
identity~\cite[Appendix, Eqs.~(54)--(55)]{EmersonAlickiZyczkowski2005}
gives
\begin{equation}
\label{eq:dec:twirled-projection}
\E_U\mathcal P_U
=\mathcal N+t(\operatorname{id}_{AE}-\mathcal N),
\qquad
t=\frac{c^2-1}{d^2-1},
\qquad
\gamma=k^2t.
\end{equation}
Indeed, the map $X\mapsto k^{-1}\Tr_D(X)\otimes\id_D$ on
$\mathcal B(A)$ is a projection of rank $c^2$. Taking its
superoperator trace gives $1+t(d^2-1)=c^2$. Tensoring with the
identity map on $E$ leaves $t$ unchanged.
Here $t>0$ and $k^2t=(1-c^{-2})/(1-d^{-2})\in[3/4,1]$.
For each $U$, let
\begin{align*}
T_U=\{\rho_{CE}^{U}-\pi_C\otimes\rho_E>0\},
\qquad
Q_U=U^*(T_U\otimes\id_D)U,
\end{align*}
and set
\begin{align*}
F=\E_UQ_U,\qquad S=F-\mathcal N(F).
\end{align*}
Since $Q_U$ is a projection, $0\leq F\leq\id_{AE}$.
The map $\mathcal N$ is positive and unital, so
$0\leq\mathcal N(F)\leq\id_{AE}$.
Hence $S$ is self-adjoint,
$-\id\leq S\leq\id$, and $\mathcal N(S)=0$.
Moreover, $\mathcal P_U(Q_U)=Q_U$. Since $\mathcal N$ is self-adjoint
for the trace pairing and $\mathcal N(\rho_{AE})=\sigma$, we have
\begin{align}
\Tr(\rho_{AE}S)
=\Tr[(\rho_{AE}-\sigma)F]=\E_U\Tr[(\rho_{CE}^{U}-\pi_C\otimes\rho_E)T_U]
=\Delta.
\label{eq:dec:strong-signal}
\end{align}

We next show that
\begin{equation}\label{eq:dec:bessel}
\mathcal N(S^2)\leq t\,\mathcal N(F).
\end{equation}
Let $\eta_E$ be any state and set $\nu=\pi_A\otimes\eta_E$ and
$w=\Tr(\nu S^2)$. Since $\mathcal N(S)=0$ and $F=S+\mathcal N(F)$,
the trace-preserving conditional expectation $\mathcal N$ gives
$w=\Tr(\nu FS)$. The operators $\nu$ and $Q_U$ belong to the range of
$\mathcal P_U$. By its bimodule identity and trace preservation,
\begin{align*}
w&=\E_U\Tr[\nu Q_U\mathcal P_U(S)],\\
\E_U\Tr[\nu\mathcal P_U(S)^2]
&=\E_U\Tr[\nu S\mathcal P_U(S)]=tw,
\end{align*}
where the last equality uses $\E_U\mathcal P_U(S)=tS$.
Cauchy--Schwarz for the Hilbert--Schmidt inner product averaged over $U$
gives
\[
w^2\leq\bigl(\E_U\Tr[\nu Q_U^2]\bigr)
          \bigl(\E_U\Tr[\nu\mathcal P_U(S)^2]\bigr)
=t\,\Tr(\nu F)\,w.
\]
Thus $w\leq t\Tr(\nu F)$, including when $w=0$. This proves
the operator inequality \eqref{eq:dec:bessel}. Indeed,
$\Tr(\nu S^2)=\Tr[\nu\mathcal N(S^2)]$ and
$\Tr(\nu F)=\Tr[\nu\mathcal N(F)]$ for every $\eta_E$, and
both $\mathcal N(S^2)$ and $\mathcal N(F)$ act only on $E$.

The reference term in the error of the test $T_U$ gives
\[
\Tr(\sigma F)=\E_U\Tr[(\pi_C\otimes\rho_E)T_U]\leq 1-\Delta.
\]
Since $\mathcal N(\sigma)=\sigma$,
\eqref{eq:dec:bessel} and $k^2t\leq1$ imply
$k^2\Tr(\sigma S^2)\leq1-\Delta$. Together with
\eqref{eq:dec:strong-signal} and $-\id\leq S\leq\id$, this verifies the
hypotheses of Lemma~\ref{lem:test-transfer}, which gives
\eqref{eq:dec:strong-reverse}.

Apply \eqref{eq:dec:strong-reverse} at blocklength $n$ with dimensions
$c^n$ and $k^n$. The support inclusion
$\supp\rho_{AE}\subseteq\supp(\pi_A\otimes\rho_E)$ lets us use
Lemma~\ref{lem:weighted-testing} with reference $\sigma=\pi_A\otimes\rho_E$
and weight $k^{2n}$. Since
$D_s(\rho_{AE}\|\sigma)-2\log k=\log(c/k)-H_s(A|E)_\rho$ and
the factor $1/3$ does not affect the logarithmic limit, we obtain
\begin{align*}
\limsup_{n\to\infty}-\frac1n\log(1-\Delta_n)
&\leq\lim_{n\to\infty}-\frac1n\log\Tr\bigl[
\rho_{AE}^{\otimes n}\wedge k^{2n}\sigma^{\otimes n}\bigr]\\
&=\max_{0\leq s\leq1}(1-s)
\bigl(\log(c/k)-H_s(A|E)_\rho\bigr).
\end{align*}
The bound \eqref{eq:dec:known-strong} gives the matching lower bound on
$\liminf_{n\to\infty}-n^{-1}\log(1-\Delta_n)$ by additivity of
$H_s(A|E)_\rho$ and continuity at the endpoints. Hence the limit exists
and equals \eqref{eq:dec:strong}.
\end{proof}

\subsection{The reliability function}
We first control the moments of the randomized partial trace. The proof
compares Haar and Gaussian moments and applies Rosenthal's inequality
to independent Gaussian blocks.
\begin{lemma}
\label{lem:dec:moment}
Let $A=C\otimes D$ with $c=|C|\geq2$, $k=|D|\geq1$, and $d=ck$.
Suppose that $0\leq Q\leq\id_{AE}$ and $Q_E:=\Tr_A Q\leq db\id_E$ for
some $0<b\leq1$.
Set
\[
Q_0=Q-\pi_A\otimes Q_E,
\qquad
K_U=k\Tr_D(UQ_0U^*).
\]
Then, for every $p\geq2$,
\begin{equation}
\label{eq:dec:moment-bound}
\left(\E_U\Tr|K_U|^p\right)^{1/p}
\leq C_p(kd|E|)^{1/p}\max\{1,k\sqrt b\},
\end{equation}
where $C_p=6K_pK_{2p}^2$ and $K_p\geq1$ is the constant in
Lemma~\ref{lem:rosenthal-moments}.
\end{lemma}

\begin{proof}
\emph{Gaussian moment estimate.}
For matrices $B_j$, independent standard complex Gaussian variables
$g_j$ with $\E|g_j|^2=1$, and $q>2$, we first show
\begin{equation}
\label{eq:dec:gaussian-rosenthal}
\left(\E\left\|\sum_jg_jB_j\right\|_q^q\right)^{1/q}
\leq K_q\max\left\{
\left\|\left(\sum_jB_jB_j^*\right)^{1/2}\right\|_q,
\left\|\left(\sum_jB_j^*B_j\right)^{1/2}\right\|_q
\right\}.
\end{equation}
Fix $q>2$ and an integer $N\geq1$.
Let $\{g_{j\ell}\}_{j,\ell}$ be independent standard complex
Gaussian variables, and define
\[
\widetilde{B}_{j\ell}
=
\frac1{\sqrt N}
\begin{pmatrix}
0 & g_{j\ell}B_j\\
\overline{g_{j\ell}}B_j^* & 0
\end{pmatrix},\qquad
\left|\widetilde B_{j\ell}\right|
=
\frac{|g_{j\ell}|}{\sqrt N}
\begin{pmatrix}
(B_jB_j^*)^{1/2}&0\\
0&(B_j^*B_j)^{1/2}
\end{pmatrix}.
\]
These matrices are independent, mean-zero, and self-adjoint.
Since $\E|g_{j\ell}|^2=1$, their second moments satisfy
\[
\sum_j\sum_{\ell=1}^N \E \widetilde{B}_{j\ell}^2
=
\begin{pmatrix}
\sum_j B_jB_j^* & 0\\
0 & \sum_j B_j^*B_j
\end{pmatrix}.
\]
Moreover, direct computation gives
\[
\sum_j\sum_{\ell=1}^N \E\Tr|\widetilde{B}_{j\ell}|^q
=
2N^{1-q/2}\E|g_1|^q
\sum_j\|B_j\|_q^q.
\]

For each $j$, the variable
$g_j=N^{-1/2}\sum_{\ell=1}^N g_{j\ell}$
is again standard complex Gaussian, and these variables
are independent as $j$ varies. It follows that
\[
\E\Tr\left|\sum_j\sum_{\ell=1}^N \widetilde{B}_{j\ell}\right|^q
=
2\E\left\|\sum_jg_jB_j\right\|_q^q.
\]
Applying Lemma~\ref{lem:rosenthal-moments} to the family
$\{\widetilde{B}_{j\ell}\}_{j,\ell}$ and dividing by $2^{1/q}$ gives
\begin{align*}
\left(\E\left\|\sum_jg_jB_j\right\|_q^q\right)^{1/q}
\leq K_q\max\Biggl\{
&N^{1/q-1/2}
\left(\E|g_1|^q\sum_j\|B_j\|_q^q\right)^{1/q},
\\
&\left[
\frac12\Tr\left(\sum_jB_jB_j^*\right)^{q/2}
+
\frac12\Tr\left(\sum_jB_j^*B_j\right)^{q/2}
\right]^{1/q}
\Biggr\}.
\end{align*}
Since $q>2$, the first term tends to zero as $N\to\infty$,
whereas the second term is independent of $N$.
The latter is at most
\[
\max\left\{
\left\|\left(\sum_jB_jB_j^*\right)^{1/2}\right\|_q,
\left\|\left(\sum_jB_j^*B_j\right)^{1/2}\right\|_q
\right\}.
\]
Letting $N\to\infty$ therefore proves
\eqref{eq:dec:gaussian-rosenthal}.

\emph{Comparison of Haar and Gaussian moments.}
Let $G$ be a $d\times d$ matrix with independent circularly symmetric
complex Gaussian entries of variance $1/d$.
In the polar decomposition $G=UR$, the unitary $U$ is Haar
distributed and independent of $R=(G^*G)^{1/2}$.
Unitary invariance and $\E_R R^2=\id_A$ give
\begin{equation}
\label{eq:dec:polar-map}
\E_R RXR
=\lambda_d X+(1-\lambda_d)\pi_A\otimes\Tr_A X,
\qquad
\lambda_d=
\frac{\E[(\Tr R)^2]-1}{d^2-1}.
\end{equation}
Here and below, matrices acting on $A$ act trivially on $E$.
The expression for $\lambda_d$ follows by taking the
superoperator trace, since the map $X\mapsto RXR$ has
superoperator trace $(\Tr R)^2$ on $\mathcal B(A)$.
Gaussian second and fourth moments give
\[
\E\Tr R^2=d,
\qquad
\E\Tr R^4=2d.
\]
H\"older's inequality yields
\[
d\leq(\E\Tr R)^{2/3}(\E\Tr R^4)^{1/3},
\]
so $\E\Tr R\geq d/\sqrt2$. Hence
\[
\lambda_d\geq\frac{d^2/2-1}{d^2-1}\geq\frac13.
\]
Since $\Tr_A Q_0=0$, conditional Jensen's inequality in
\eqref{eq:dec:polar-map} implies
\begin{equation}
\label{eq:dec:haar-gaussian}
\left(\E_U\Tr|K_U|^p\right)^{1/p}
\leq3\left(
\E_G\left\|k\Tr_D(GQ_0G^*)\right\|_p^p
\right)^{1/p}.
\end{equation}

\emph{Independent Gaussian blocks.}
For $z=1,\ldots,k$, put
\[
S_z=\sqrt k\,(\id_C\otimes\langle z|_D)G,
\qquad
Z_z=(S_z\otimes\id_E)Q_0(S_z^*\otimes\id_E).
\]
The matrices $S_z$ are independent $c\times d$ Gaussian
matrices with entry variance $1/c$, and
\[
k\Tr_D(GQ_0G^*)=\sum_{z=1}^k Z_z.
\]
Gaussian second and fourth moments, together with
$\Tr_A Q_0=0$, give
\begin{align*}
\E Z_z&=0,\\
\E Z_z^2
&=\frac{\id_C}{c}\otimes\Tr_A Q_0^2=\frac{\id_C}{c}\otimes
\left(\Tr_A Q^2-\frac1dQ_E^2\right)
\leq kb\id_{CE}.
\end{align*}

To bound the $p$-th moment of each $Z_z$, take
$P=Q$ or $P=\pi_A\otimes Q_E$.
In both cases,
$0\leq P\leq\id_{AE}$ and $\Tr_A P=Q_E\leq db\id_E$.
Writing the entries of $S_z$ as $g_{ui}/\sqrt c$, we have
\[
(S_z\otimes\id_E)P^{1/2}
=\sum_{u=1}^c\sum_{i=1}^d g_{ui}B_{ui},
\qquad
B_{ui}=
\frac1{\sqrt c}(|u\rangle\langle i|\otimes\id_E)P^{1/2}.
\]
The two square functions satisfy
\[
\sum_{u,i}B_{ui}^*B_{ui}=P,
\qquad
\sum_{u,i}B_{ui}B_{ui}^*
=\frac{\id_C}{c}\otimes\Tr_A P.
\]
Applying \eqref{eq:dec:gaussian-rosenthal} with $q=2p$
and then squaring gives
\begin{align*}
\left(
\E\left\|(S_z\otimes\id_E)P(S_z^*\otimes\id_E)\right\|_p^p
\right)^{1/p}
\leq&
K_{2p}^2\max\left\{
\|P\|_p,
\left\|\frac{\id_C}{c}\otimes\Tr_A P\right\|_p
\right\}\\
\leq&
K_{2p}^2(d|E|)^{1/p}\max\{1,kb\}.
\end{align*}
The triangle inequality therefore implies
\[
\left(\E\Tr|Z_z|^p\right)^{1/p}
\leq2K_{2p}^2(d|E|)^{1/p}\max\{1,kb\}.
\]

Apply Lemma~\ref{lem:rosenthal-moments} to the independent,
mean-zero, self-adjoint matrices $Z_1,\ldots,Z_k$.
The preceding estimates yield
\begin{align*}
\left(\E\Tr\left|\sum_{z=1}^kZ_z\right|^p\right)^{1/p}
&\leq K_p\max\left\{
2K_{2p}^2(kd|E|)^{1/p}\max\{1,kb\},
(c|E|)^{1/p}k\sqrt b
\right\}\\
&\leq
2K_pK_{2p}^2(kd|E|)^{1/p}\max\{1,k\sqrt b\}.
\end{align*}
The last step uses $b\leq\sqrt b$, $c\leq kd$, and
$K_{2p}\geq1$.
Combining this with \eqref{eq:dec:haar-gaussian} proves
\eqref{eq:dec:moment-bound}.
\end{proof}

We now combine the moment estimate with the Haar-average identity
\eqref{eq:dec:twirled-projection} to obtain the trace-distance lower
bound needed for reliability.

\begin{lemma}
\label{lem:dec:testing-lower}
Let $0<b\leq1$, and suppose that $0\leq T\leq\id_{AE}$ satisfies
\[
\Tr_A[(\pi_A\otimes\id_E)T]\leq b\id_E.
\]
Then, for every $p\geq2$,
\begin{equation}
\label{eq:dec:one-shot-reverse}
\frac12\E_U
\left\|\rho_{CE}^U-\pi_C\otimes\rho_E\right\|_1
\geq\frac12
\left(
\frac{\gamma[\Tr(\rho_{AE}T)-b]_+}
{C_p(kd|E|)^{1/p}\max\{1,k\sqrt b\}}
\right)^{p/(p-1)},
\end{equation}
where $C_p$ is the constant in Lemma~\ref{lem:dec:moment}.
\end{lemma}
\begin{proof}
Use Lemma~\ref{general-one-shot} with
$\mathcal M=L^\infty(\mathcal U(A),\mu_{\mathrm{Haar}})\bar\otimes\mathcal B(CE)$,
and let $\mathcal E=\E_U$ be the conditional expectation onto the
constant $\mathcal B(CE)$-valued functions. The trace is
$\Tr_{\mathcal M}=\E_U\Tr$. Take
\[
\varphi=\frac{\id_{CE}}{|CE|},\qquad
\rho(U)=\rho_{CE}^U,\qquad
S(U)=k\Tr_D\bigl[U(T-\mathcal N(T))U^*\bigr].
\]
Haar averaging gives $\mathcal E(\rho)=\pi_C\otimes\rho_E$ and
$\mathcal E(S)=0$. Since both $\rho(U)$ and $\mathcal E(\rho)$ are states,
$-\id_{CE}\leq\rho(U)-\mathcal E(\rho)\leq\id_{CE}$, so the domination
hypothesis holds with $L=|CE|$. Write
$\Delta=\frac12\E_U\norm{\rho_{CE}^U-\pi_C\otimes\rho_E}_1$.
Lemma~\ref{general-one-shot} yields
\begin{equation*}
\bigl|\Tr_{\mathcal M}(\rho S)\bigr|
\leq (2\Delta)^{1-1/p}|CE|^{1/p}\|S\|_{p,\varphi}
=(2\Delta)^{1-1/p}\bigl(\E_U\Tr|S(U)|^p\bigr)^{1/p}.
\end{equation*}

By \eqref{eq:dec:twirled-projection} and the definition of
$\mathcal P_U$,
\begin{align*}
\Tr_{\mathcal M}(\rho S)=k^2\Tr\bigl[\rho_{AE}(\E_U\mathcal P_U)(T-\mathcal N(T))\bigr]=\gamma\left(\Tr(\rho_{AE}T)-\frac1d\Tr(\rho_E\Tr_A T)\right).
\end{align*}
The test constraint therefore gives
$|\Tr_{\mathcal M}(\rho S)|\geq\gamma[\Tr(\rho_{AE}T)-b]_+$.
Lemma~\ref{lem:dec:moment} with $Q=T$ bounds the moment on the right by
$C_p(kd|E|)^{1/p}\max\{1,k\sqrt b\}$. Combining these bounds and
rearranging proves \eqref{eq:dec:one-shot-reverse}.
\end{proof}

\begin{proof}[Proof of \eqref{eq:dec:reliability}]
Assume $\rho_{AE}\ne\pi_A\otimes\rho_E$, so that
\[
D_1^*(\rho_{AE}\|\pi_A)
=D(\rho_{AE}\|\pi_A\otimes\rho_E)>0.
\]
The direct bound \eqref{eq:dec:known-error} and additivity of
$H_\alpha^*(A|E)_\rho$ give
\[
\liminf_{n\to\infty}-\frac1n\log\Delta_n
\geq
\max_{1\leq\alpha\leq2}\frac{\alpha-1}{\alpha}
\left(H_\alpha^*(A|E)_\rho-\log\frac ck\right),
\]
where the endpoint values are included by continuity.

For the matching upper bound on the reliability exponent, we apply
Lemma~\ref{lem:dec:testing-lower} to an optimal conditional test.
The testing pair in Theorem~\ref{thm:best-testing-exponent} is
$(\rho_{AE},\pi_A)$; write $A_n(r)$ for its optimal success
probability.
Fix $r>0$. The feasible set defining $A_n(r)$ is compact,
so we may choose a maximizing test $T_n$ satisfying
\[
\Tr(\rho_{AE}^{\otimes n}T_n)=A_n(r),
\qquad
\frac1{d^n}\Tr_{A^n}T_n\leq e^{-nr}\id_{E^n}.
\]
For every fixed $p\geq2$, apply
Lemma~\ref{lem:dec:testing-lower} at blocklength $n$ with
$b=e^{-nr}$.
Since the corresponding Haar coefficient is at least $3/4$,
\begin{equation}
\label{eq:dec:testing-profile}
\Delta_n\geq\frac12
\left(
\frac{3[A_n(r)-e^{-nr}]_+}
{4C_p(kd|E|)^{n/p}\max\{1,k^ne^{-nr/2}\}}
\right)^{p/(p-1)}.
\end{equation}
Monotonicity of the sandwiched R\'enyi divergence in its order
implies
\begin{equation}\label{eq:p-d-1}
\mathsf H_r^*(\rho_{AE}\|\pi_A)
\leq[r-D_1^*(\rho_{AE}\|\pi_A)]_+<r.
\end{equation}
By Theorem~\ref{thm:best-testing-exponent}, we have
\begin{equation}\label{eq:p-d-2}
A_n(r)=\exp[-n(\mathsf H_r^*(\rho_{AE}\|\pi_A)+o(1))].
\end{equation}
Combining \eqref{eq:p-d-1} and \eqref{eq:p-d-2} yields
\[
\lim_{n\to\infty}-\frac1n\log[A_n(r)-e^{-nr}]_+
=\mathsf H_r^*(\rho_{AE}\|\pi_A).
\]
For fixed $p$, we also have
\[
\frac1n\log\max\{1,k^ne^{-nr/2}\}
=\frac12[2\log k-r]_+.
\]
Taking negative logarithms in \eqref{eq:dec:testing-profile}, dividing
by $n$, and letting $n\to\infty$ with $p$ fixed gives
\[
\limsup_{n\to\infty}-\frac1n\log\Delta_n
\leq\frac p{p-1}
\left[
\mathsf H_r^*(\rho_{AE}\|\pi_A)
+\frac12[2\log k-r]_+
+\frac{\log(kd|E|)}p
\right],
\]
where the term $n^{-1}\log(4C_p/3)$ vanishes because
$C_p$ depends only on $p$.
Letting $p\to\infty$ yields
\begin{equation}
\label{eq:dec:testing-exponent-bound}
\limsup_{n\to\infty}-\frac1n\log\Delta_n
\leq \mathsf H_r^*(\rho_{AE}\|\pi_A)
+\frac12[2\log k-r]_+,
\qquad r>0.
\end{equation}

Taking the infimum over $r>0$ in
\eqref{eq:dec:testing-exponent-bound} and applying
\eqref{eq:scalar-testing-identity} with $\ell=2\log k$ gives
\begin{align*}
\limsup_{n\to\infty}-\frac1n\log\Delta_n
&\leq\max_{1\leq\alpha\leq2}\frac{\alpha-1}{\alpha}
\bigl(2\log k-D_\alpha^*(\rho_{AE}\|\pi_A)\bigr)\\
&=\max_{1\leq\alpha\leq2}\frac{\alpha-1}{\alpha}
\bigl(H_\alpha^*(A|E)_\rho-\log(c/k)\bigr),
\end{align*}
where the equality uses $D_\alpha^*(\rho_{AE}\|\pi_A)=\log d-H_\alpha^*(A|E)_\rho$.
Together with the direct bound, this proves that the limit
exists and establishes \eqref{eq:dec:reliability}.
\end{proof}

%%%%%%%%%%%%%%%%%%%%%%%%%%%%%%%%%%%%%%%%%%%%%%%%%%%%%%%%

\medskip
\textbf{Acknowledgments.} --- 
H.C.~acknowledges the support by NSTC 115-2628-E-002-005, NSTC 114-2119-M-001-002, and NSTC 115-2124-M-002-014, NTU-115V2016-1, NTU-CC115L893705, and the Yushan Fellow Program by the Ministry of Education (MOE) MOE-114-YSFEE-0003-006-P2. 

S.L. is supported by NSFC under Grant 62571166. H.Q. is supported by the National Natural Science Foundation
of China (Nos. 12371138 and W2441002).

L.G. is partially supported by the National Natural Science
Foundation of China (grant No.~12401163) and the Department of Science
and Technology of Hubei Province (project Nos.~2025EHA041 and
2025AFA044). 
L.G. is grateful to Marius Junge for pointing out the potential applications of the noncommutative Rosenthal inequality to the exact exponent for convex splitting back in 2024.

\medskip
\textbf{Statement on use of AI.} --- 
We use ChatGPT-6 Astra to execute our ideas of using Rosenthal inequalities to obtain the exponents.
Part of the paragraphs are also generated by ChatGPT-6 Astra.
The authors made edition and verified the derivations.

\medskip
\textbf{Note added.} --- See also the concurrent work~\cite{concurrent} by Bartosz Regula and Hao-Chung Cheng for related results on the trace-distance exponents for privacy amplification.  

\bibliographystyle{IEEEtran}
\bibliography{references}
\end{document}